\documentclass{jpp}

\usepackage{graphicx}
\usepackage{verbatim}
\usepackage{amsmath}
\usepackage{amssymb}
\usepackage[loading]{tracefnt}

\newcommand{\gyav}[1]{\left\langle $1 \right\rangle_\mathbf{R}}
\newcommand{\vdrift}[2][s]{\ensuremath{\mathbf{v_{d}}_{s}}}
\usepackage{natbib}

\begin{document}
%\title[Gyrokinetic transport properties of alpha particles in turbulence]{Gyrokinetic transport properties of non-Maxwellian alpha particles in electrostatic turbulence}
%\title[Assumptions on modelling alpha particles in gyrokinetics]{Analysis of common assumptions in the modelling of alpha particles in electrostatic gyrokinetics}
\title[Alpha particles in gyrokinetics]{Validating modelling assumptions of alpha particles in electrostatic turbulence}
\author[G. Wilkie, I. Abel, E. Highcock, W. Dorland]{G. Wilkie$^1$ 
  \thanks{Email address for correspondence: gwilkie@umd.edu} 
 , I. Abel$^2$, E. Highcock$^3$, W. Dorland$^1$ }
%\ead{gwilkie@umd.edu}

%\affiliation{$^1$ Institute for Research in Electronics and Applied Physics, University of Maryland, College Park, MD 20742, USA\\ [\affilskip]
%$^2$ Princeton Center for Theoretical Science, Princeton University, Princeton, NJ 08544, USA \\ [\affilskip]
%$^3$ Rudolph Peierls Centre for Theoretical Physics, University of Oxford, Oxford OX1 3NP, UK }
\affiliation{$^1$ University of Maryland, College Park, MD 20742, USA\\ [\affilskip]
$^2$ Princeton University, Princeton, NJ 08544, USA \\ [\affilskip]
$^3$ University of Oxford, Oxford OX1 3NP, UK }

\maketitle
\begin{abstract}
To rigorously model fast ions in fusion plasmas, a non-Maxwellian equilibrium distribution must be used. In the work, the response of high-energy alpha particles to electrostatic turbulence has been analyzed for several different tokamak parameters. Our results are consistent with known scalings and experimental evidence that alpha particles are generally well-confined: on the order of several seconds. It is also confirmed that the effect of alphas on the turbulence is negligible at realistically low concentrations, consistent with linear theory. It is demonstrated that the usual practice of using a high-temperature Maxwellian gives incorrect estimates for the radial alpha particle flux, and a method of correcting it is provided. Furthermore, we see that the timescales associated with collisions and transport compete at moderate energies, calling into question the assumption that alpha particles remain confined to a flux surface that is used in the derivation of the slowing-down distribution.

\end{abstract}

\section{Introduction}

The study of fusion-related plasmas is intimately concerned with the behavior of alpha particles, which initially carry an energy of $E_\alpha \equiv 3.5 \mathrm{MeV}$ as a product of the deuterium-tritium (DT) nuclear fusion reaction, which is much faster than the $\sim10$ keV ions that make up the plasma bulk. The high-energy alpha particles give up their energy primarily by colliding against electrons, eventually forming a high-energy tail in collisional equilibrium \citep[see][]{Gaffey1976}. 

How well this non-Maxwellian population of high-energy particles is confined is a critical question for the possibility of achieving ignition. It is then no surprise that there has been a considerable amount of work done on the topic. \citet{Estrada-Mila2006} performed numerical simulations using \texttt{GYRO} \citep{Candy2003} and found significant transport of high-energy alpha particles in the core due to electrostatic turbulence. This was confirmed by \citet{Albergante2009} using \texttt{GENE} \citep{Jenko2000}, and it was stressed that turbulence can result in the retention of low-energy Helium ash, a result which \citet{Angioni2009} also found with \texttt{GS2} \citep{Kotschenreuther1995,Dorland2000}. All of these nonlinear gyrokinetic simulations of alpha particle transport were performed by treating the alpha particle population as a hot Maxwellian species, using the so-called ``equivalent Maxwellian'' approximation. This approximation has been in use for a long time \citep[e.g.][]{Rosenbluth1975}, and was formalised by \citet{Estrada-Mila2006}.

However, alpha particles are not Maxwellian in reality \citep{Gaffey1976}, a fact which is well-recognised in the above references. Indeed, care was taken to show that using a Maxwellian of the same temperature gave accurate linear results. Then, quasilinear simulations in \texttt{GS2} were performed using the non-Maxwellian slowing-down distribution \citep{Angioni2008} to obtain the radial flux of alphas particles, where attention was drawn to the incorrectness of the radial gradient of the equivalent Maxwellian. While it was found that good estimates for the diffusion coefficient can be obtained using the Maxwellian approximation, in this work we demonstrate that the numerical value of the alpha particle flux is in fact poorly estimated, depending on the local parameters used. Even when the equivalent Maxwellian is inadequate, we present a method to rigorously obtain the correct energy-dependent flux valid in the trace limit.

By treating the electrostatic ion-scale turbulence as a given background field with known properties, to which the energetic particles passively react, analytic scalings can be obtained. It was found \citep{Hauff2009, Hauff2009a} that the diffusion of energetic particles scales inversely with energy ($E^{-1}$) for particles with high pitch angle ($v_\| \sim v$), and as $E^{-3/2}$ for deeply-trapped energetic particles. It was later pointed out \citep{Pueschel2012} that this is an expansion in Larmor radius, with the former result valid only for prohibitively large pitch angles for high-energy particles. Therefore, an overall $E^{-3/2}$ scaling is expected: a result we confirm.

In this work, we present a fully nonlinear self-consistent treatment of non-Maxwellian energetic particles. We use this capability to test commonly-made assumptions in the modelling of alpha particles in the context of turbulence. After reviewing our approach to the problem in section 2, we will find in section 3 that the passive-tracer limit is largely satisfactory. However, the equivalent-Maxwellian approach is not adequate in determining the transport properties of alpha particles (see section 4). Later, in section 5, we estimate how well alpha particles of various energies are confined for realistic equilibrium parameters, followed by a general discussion of these results. 

\section{Background: Gyrokinetics of fast ions}

Here we give a brief exposition of gyrokinetics with an eye to non-Maxwellian energetic particles. The full derivation of the gyrokinetic ordering for weakly collisional species is given in \citet{Abel2014}. We will assume that the equilibrium distribution function is isotropic in velocity space: that is, only a function of energy.
%For previous derivations of gyrokinetics, see for example: \citet{Antonsen1980, Frieman1982, Sugama1998, Brizard2007, Abel2013b}.

\subsection{Low-collisionality Gyrokinetics}

Gyrokinetics is the standard tool for studying low-frequency, small-scale turbulence in highly magnetised plasmas. A strong magnetic field allows one to take advantage of the strong anisotropisation of the dynamics that results, and one can perform an asymptotic expansion of the Fokker-Planck equation in the small parameter $\rho^* \equiv \rho_i/a$, where $a$ is the minor radius of the device (representing the equilibrium scale length), $\rho_i = v_{ti} / \Omega_i$ is the characteristic Larmor radius of the bulk ions (with $v_{ts} \equiv \sqrt{2 T_s / m_s}$ the thermal speed of species $s$ with temperature $T_s$ and mass $m_s$), and the gyrofrequency is $\Omega_s = Z_s e B / m_s c$. The equilibrium magnetic field $\mathbf{B}$, temperatures $T_s$, and densities $n_s$ vary on the scale of $a$, as do any fluctuating quantities along the magnetic field, characterized by the parallel wavenumber $k_\|$. The advantage of this expansion is that it allows an averaging over the fast gyro-motion while retaining fine-spatial-scale dynamics perpendicular to the field, characterised by the wavenumber $k_\perp$, the scale of which is allowed to be as small as the Larmor radius.   To summarise, the gyrokinetic ordering is such that:
\begin{equation}\label{gkorderings}
\frac{|\mathbf{v_E}|}{v_{ti}} \sim \frac{k_\|}{k_\perp} \sim \frac{\omega}{\Omega_i} \sim \frac{\rho_i}{a} \equiv \rho^*,
\end{equation}
where $\omega$ is a characteristic frequency associated with turbulent fluctuations in the distribution function, and $\mathbf{v_E}$ is the $E \times B$ drift velocity. Contrast this with, for example, drift-kinetics or magnetohydrodynamics, both of which require $k_\perp \ll 1/\rho_i$, but allows $v_E \sim v_{ti}$ or $\omega \sim \Omega_i$  respectively. 

The distribution function is decomposed as 
\begin{equation}
f_s\left(\mathbf{r},E,\mu,\xi,\sigma_\|,t \right) = F_{0s} + \delta f_s = F_{0s} + Z_s e \phi \frac{\partial F_{0s}}{\partial E} + h_s,
\end{equation}
and in general depends on spatial position $\mathbf{r}$ through the electrostatic potential $\phi(\mathbf{r},t)$. In these coordinates, the sign of $v_\|$ must be specified by $\sigma_\|$ so that $v_\| = \sigma_\| \sqrt{\left(2/m_s\right)\left(E - \mu B\right)}$ and $v_\perp = \sqrt{2 \mu B/m_s }$. The direction of $\mathbf{v_\perp}$ is determined from the gyro-phase $\xi$. The gyro-center position $\mathbf{R}_s$ is related to $\mathbf{r}$ by $\mathbf{r} = \mathbf{R}_s + \mathbf{b}\times \mathbf{v_\perp}/\Omega_s$. The slowly-evolving equilibrium distribution is $F_{0s}$, and is written this way when no particular velocity dependence is specified. We will introduce notations such as $F_{Ms}$ and $F_{Ss}$ later, and these shall be interpreted as specific forms of $F_{0s}$ with given velocity dependence. The non-adiabatic part of the perturbed distribution, $h_s \sim \rho^* F_{0s}$ is a function of gyro-center position and velocity, but does not depend on gyro-phase $\xi$. It is found by solving the gyrokinetic equation, which in the electrostatic ($\beta \rightarrow 0$) limit without equilibrium flow reads:
\begin{align}
\frac{\partial h_s}{\partial t} &+ \left( v_\| \mathbf{b} + \vdrift + \frac{c}{B} \mathbf{b} \times \nabla \left\langle \phi \right\rangle_{\mathbf{R}_s} \right) \cdot \nabla h_s - C_{GK}\left[h_s\right] - Z_s e \frac{\partial \left\langle \phi \right\rangle_{\mathbf{R}_s}}{\partial t} \frac{\partial h_s}{\partial E} \label{lowcollgk}\\
&= - Z_s e \frac{\partial F_{0s}}{\partial E} \frac{\partial \left\langle \phi \right\rangle_{\mathbf{R}_s}}{\partial t} - \frac{c}{B} \mathbf{b} \times \nabla \left\langle \phi \right\rangle_{\mathbf{R}_s} \cdot \nabla F_{0s}, \nonumber
\end{align}
where $C_{GK}$ is an appropriate gyro-averaged collision operator \citep[see][]{Abel2008, Barnes2009a,Li2011}, $Z_s e$ is the charge carried by species $s$, and $\mathbf{v_d}_s$ is the velocity of magnetic drifts due to the curvature and gradient of $\mathbf{B}$. All gradients in equation (\ref{lowcollgk}) are with respect to the gyrocenter coordinate $\mathbf{R}_s$. 

Equation (\ref{lowcollgk}) is closed by solving for the electrostatic potential via the quasineutrality condition: 
\begin{equation}\label{qneutrality}
\phi \sum\limits_s Z_s^2 e^2 \int \frac{\partial F_{0s}}{\partial E} d^3 \mathbf{v} + \sum\limits_s Z_s e \int \left\langle h_s \right\rangle_\mathbf{r} d^3 \mathbf{v} = 0,
\end{equation}
in which $\left\langle \right\rangle_\mathbf{r}$ is the gyro-average operation at constant spatial position $\mathbf{r}$. 

The form of equation (\ref{lowcollgk}) is identical to other iterative derivations of the gyrokinetic equation \citep[see][]{Frieman1982,Sugama1998,Abel2013b}, except for the final term on the left-hand side. This is the so-called parallel nonlinearity as expressed in $E$, $\mu$ coordinates (it's name comes from the form it takes in $v_\|$, $\mu$ coordinates). Its presence at this order is a consequence of the low-collisionality ordering because now, much finer velocity-space structures can develop in $h_\alpha$ such that $\partial h_\alpha / \partial E \sim \left(\rho^{*}\right)^{-1} h_\alpha / E$. Hamiltonaian derivations of the gyrokinetic equation \citet{Brizard2007} also include this nonlinearity, but for different reasons (e.g. its convenient conservation properties, see the appendix of \citet{Abel2013b}). In order to properly resolve this term, one needs a velocity-space grid $\left( 1 / \rho^* \right)$ finer than usual, making its inclusion numerically challenging. It is important to note that this term is only included in the gyrokinetic equation for the alpha particles; the distribution functions, $h_s$, for the bulk (non-energetic) species are obtained by solving the standard gyrokinetic equation with $F_{0s}$ Maxwellian.

At higher order in $\rho^{*}$, one obtains the transport equation, which now contains the collision operator acting on $F_{0s}$. After integrating over $\mu$ and summing over the sign of the parallel velocity, but preserving the energy dependence, the transport equation becomes:
\begin{equation}
\label{transporteqn}
\frac{1}{V'} \frac{\partial}{\partial t} V' F_{0s} + \frac{1}{V'} \frac{\partial}{\partial \psi} V' \Gamma_s(E) + \frac{1}{\sqrt{E}} \frac{\partial}{\partial E} \sqrt{E}\Gamma_{E,s}(E) = \left\langle   C\left[F_{0s}\right] + S_s \right\rangle_\psi.
\end{equation}
The source of particles of species $s$ is denoted as $S_s$.
Equation \ref{transporteqn} determines the slow-time evolution of the equilibrium, including the average transport of particles in phase space due to turbulence. The radial flux is:
\begin{equation}\label{fluxedef}
\Gamma_s\left(E\right) \equiv \left\langle \sum\limits_{\sigma_\|} \int  h_s\left\langle  \mathbf{v_E} \right\rangle_{\mathbf{R}_s} \cdot\nabla\psi \, \frac{\pi B \mathrm{d} \lambda }{ \sqrt{1 - \lambda B}} \right\rangle_{t,\psi},
\end{equation}
and the flux in energy, representing the acceleration of particles by fluctuations, is:
\begin{equation}\label{efluxdef}
\Gamma_{E,s}\left(E\right) \equiv Z_s e \left\langle \sum\limits_{\sigma_\|} \int h_s \frac{\partial \left\langle \phi \right\rangle_{\mathbf{R}_s}}{\partial t}  \, \frac{\pi B \mathrm{d} \lambda }{ \sqrt{1 - \lambda B}} \right\rangle_{t,\psi}.
\end{equation}
In these expressions, the coordinate $\lambda \equiv \mu / E$ is used here for convenience so that the velocity space volume element is separable. Also, 
\begin{equation}
V'(\psi) \equiv \lim\limits_{\delta \psi \rightarrow 0} \frac{1}{\delta \psi} \int\limits_{\Delta} d^3\mathbf{r}, 
\end{equation}
where the domain of spatial integration $\Delta$ is the toroidal annulus formed between flux surfaces $\psi$ and $\psi + \delta \psi$.
The average in the definition of the fluxes is the composition of the average $\left\langle \ldots \right\rangle_{\psi}$ over a flux tube at the radial location $\psi$ and the time average
$\left\langle \ldots \right\rangle_t \equiv \tfrac{1}{T} \int^{t_0 + T}_{t_0}\ldots dt$ over a time $T$. The width of the flux tube is taken to be several turbulent correlation lengths and the time $T$ to be many turbulent correlation times.

We define the energy-integrated particle and heat fluxes of species $s$ as follows:
\begin{equation}\label{pfluxdef}
\Gamma_s = \frac{\sqrt{2}}{ m_s^{3/2}} \int \Gamma_s\left( E \right) \sqrt{E} dE,
\end{equation}
and 
\begin{equation}\label{hfluxdef}
Q_s = \frac{\sqrt{2}}{ m_s^{3/2}} \int \Gamma_s\left( E \right) E^{3/2} dE.
\end{equation}
Now, if we were to integrate (\ref{transporteqn}) multiplied by powers of $E$ over all energies we would recover the usual transport equations for density and heat with fluxes defined by (\ref{pfluxdef}) and (\ref{hfluxdef}), respectively. Equation (\ref{fluxedef}) defines the flux of particles that possess kinetic energy between $E$ and $E + \mathrm{d}E$ passing through a flux surface labelled by $\psi$. Using this definition means that if the distribution were an isotropic ``beam'' of particles with energy $E_{b} = m_s v_b^2 /2$ (i.e. $\delta f \propto \delta\left(E - E_b \right)$), then $\Gamma_\alpha \left( E \right) = \Gamma_\alpha \delta^3\left( \mathbf{v} - \mathbf{v_b} \right)$. That is, $\Gamma\left( E \right)$ represents what the particle flux would be of a beam of energy $E$.

If the collision operator is dominant in equation (\ref{transporteqn}), we find that the solution is $F_{0s}=F_{Ms}$, the Maxwellian distribution: 
\begin{equation}\label{maxwdist}
F_{Ms} \equiv n_s \left( \frac{m_s}{2 \pi T_s} \right)^{3/2} \exp\left(-\frac{m_s v^2}{2 T_s}\right),
\end{equation}
and terms such as $\partial F_{0s} / \partial E$ in equations (\ref{gkeqn}) and (\ref{qneutrality}) become $-F_{Ms} / T_s$. Knowing the form of $F_{0s}$, tools such as \texttt{Trinity} \citep{Barnes2009a} or \texttt{TGYRO} \citep{Candy2009} solve for the moments of equation (\ref{transporteqn}) (without the collision operator) to simulate the long-time global evolution of the toroidal device. 

As our low-collisionality ordering permits an arbitrary $F_{0\alpha}(\psi,E)$, one loses the easily parametrised form of $F_{0\alpha}$. Thus, picking an $F_{0\alpha}$ necessitates guessing (or, finding experimentally or numerically) a solution to the transport equation (\ref{transporteqn}). So, as a first step, let us employ further assumptions that allow us to use a known analytic form of $F_{0\alpha}$.

\subsection{Subsidiary expansion in $\tau_s / \tau_E$ for alphas}

To remedy the aforementioned complications associated with the low-collisionality ordering, for alpha particles we perform a subsidiary expansion on equations (\ref{lowcollgk}) and (\ref{transporteqn}). Define a parameter $\delta \equiv \tau_s / \tau_E$, where $\tau_E$ is the energy confinement time of the plasma bulk, representing the transport time scale, and $\tau_s$ is the slowing-down time, from the dominant collision frequency for high-energy alpha particles slowing down via drag on electrons \citep{Helander2002a}:
\begin{equation}\label{slowdowntime}
\tau_s \equiv \frac{3}{16 \sqrt{\pi}} \frac{ m_\alpha m_e v_{te}^3}{Z_\alpha^2 e^4 n_e \ln\Lambda},
\end{equation} 
where $\ln \Lambda$ is the Coulomb logarithm. If we take $\delta \ll 1$, collisions are now a bit stronger so that the energy derivative is $\partial h_\alpha / \partial E \sim \mathrm{O} \left[ \sqrt{\delta / \rho^*} \left(h_\alpha /E\right) \right]$, which results in a term we take to be small enough to leave out of the gyrokinetic equation. Now we have:
\begin{align}\label{gkeqn}
\frac{\partial h_s}{\partial t} &+ \left( v_\parallel \mathbf{b} + \vdrift + \frac{c}{B} \mathbf{b} \times \nabla \left\langle \phi \right\rangle_{\mathbf{R}_s} \right) \cdot \nabla h_s - C_{GK}\left[h_s\right] \\
&= - Z_s e \frac{\partial F_{0s}}{\partial E} \frac{\partial \left\langle \phi \right\rangle_{\mathbf{R}_s}}{\partial t} - \frac{c}{B} \mathbf{b} \times \nabla \left\langle \phi \right\rangle_{\mathbf{R}_s} \cdot \nabla F_{0s},\nonumber
\end{align}
and equation (\ref{transporteqn}) becomes:
\begin{equation}\label{CeqS}
\frac{\partial F_{0s}}{\partial t} = C \left[ F_{0s} \right] + \frac{\sigma}{4 \pi v_\alpha^2} \delta\left(v - v_\alpha\right).
\end{equation}
where $\sigma$ is the creation rate of alpha particles from fusion per unit volume per unit time. The alpha particle source is approximated here by a Dirac delta function at $v_\alpha = \sqrt{2 E_\alpha / m_\alpha}$. 
%This is an approximation to the spread of energies created in the center-of-mass frame of fusing ions (an analogous distribution for neutrons was found by \citet{Brysk1973}). 
The transport terms in equation (\ref{transporteqn}) are smaller by a factor of $\delta$ and therefore do not appear in equation (\ref{CeqS}), consistent with our ordering.

Table 1 lists some characteristic parameters in several fusion devices. From inspection, one can see that the low-collisionality ordering is required, and the subsidiary expansion in $\tau_s / \tau_E$ is marginally justified, but only for ITER. In this work, we will use the turbulent flux of alpha particles to verify this expansion and that $\tau_\Gamma \sim \tau_E$, where the former is the alpha particle transport time, which is the relevant time scale for alpha particle transport in equation (\ref{transporteqn}).  \begin{table} \caption{\label{tokamakparamstable} Properties of some typical tokamak properties, using the radial-average values from \citet{Budny2002}. Ion species is deuterium, and alpha particle parameters are taken at 3.5 MeV.  } \begin{center}
\begin{tabular}{@{}lllll}
\hline
& &TFTR&JET&ITER\\
\hline
Toroidal ion gyroradius & $\rho_i / a$ & 0.0028 & 0.0037 & 0.0013 \\
Toroidal alpha gyroradius &$\rho_\alpha / a$ & 0.040 & 0.058 & 0.018 \\
Poloidal ion gyroradius & $\rho_{i,pol} / a$ & 0.026 & 0.026 & 0.0079 \\
Poloidal alpha gyroradius & $\rho_{\alpha,pol} / a$ & 0.37 & 0.41 & 0.11 \\
Ion-ion collision frequency & $\nu_{ii} a / v_{ti}$ & $2.7\times10^{-5}$ & $5.0\times10^{-5}$ & $1.6\times10^{-4}$ \\
$\alpha$-e slowing-down frequency & $\nu_s^{\alpha e} a / v_{ti}$ & $1.3\times10^{-6}$ & $2.4\times10^{-6}$ & $5.9\times10^{-6}$\\
Slowing-down time (s) & $\tau_s$ & 0.48 & 1.0 & 0.85 \\
Energy confinement time (s) & $\tau_E$ & 0.13 & 0.59 & 2.98 \\
\end{tabular}
\end{center}
\end{table}

The reader may be justifiably concerned that the requirement that $\rho_s \ll a$ may be called into question for fast alpha particles due to their large Larmor orbits compared to the bulk ions. The Larmor radii of alpha particles is compared to the sizes of several large tokamaks in table \ref{tokamakparamstable}, where it can be seen that this approximation is in fact valid, even for newborn alphas at 3.5 MeV. A separate concern is the so-called banana-width created by the drift orbits of alpha particles due to gradients in the equilibrium magnetic field. The size of this orbit can be estimated by the poloidal Larmor radius: the Larmor radius using the poloidal magnetic field $\rho_{\alpha,pol} \equiv Z_\alpha e \mathbf{B} \cdot \mathbf{e_\theta}/ m_\alpha c$. This is in fact large, and can cause significant loss of alphas if the orbit extends to the wall of the tokamak. Such loss mechanisms are important, but beyond the scope of this work; we restrict ourselves to studying the turbulence-induced electrostatic transport of otherwise well-confined alpha particles. For similar reasons, we will also assume that equilibrium properties do not vary significantly over an alpha particle drift orbit, which for ITER is about one tenth of the minor radius for the most energetic alpha particles. This last assumption allows us to use the flux-tube approximation.

\subsection{The slowing-down distribution}

The analytic ``slowing-down distribution'' is the steady-state solution to an equation approximate to (\ref{CeqS}) that balances the collision operator with a fast particle source. If we set $\partial F_{0s} / \partial t = 0$ and approximate the collision operator in the range where $v_{ti} \ll v \ll v_{te}$, we can obtain the slowing-down distribution \citep[see][]{Gaffey1976, Helander2002a}:
\begin{equation}\label{sddef}
F_{S\alpha}\left(v\right) = \frac{3}{4 \pi \ln \left( 1 + v_c^3/v_\alpha^3 \right)} \frac{n_\alpha}{v_c^3 + v^3} H \left( v_\alpha - v \right),
\end{equation}
where $H$ is the Heaviside step function, $n_\alpha$ is the equilibrium density of alpha particles, and
\begin{equation}\label{vcdef}
v_c \equiv v_{te} \left( \frac{3 \sqrt{\pi}}{4} \sum\limits_i \frac{n_i m_e}{n_e m_i} Z_i^2 \right)^{1/3}
\end{equation}
is the critical speed. Above this speed, alpha particles primarily lose their energy via drag on faster electrons, whereas below, they primarily drag against approximately stationary ions. It is important to note that equation (\ref{sddef}) is only valid when $v \gg v_{ti}$. In fact, equation (\ref{CeqS}) lacks a steady state solution entirely, due to the particle source. The fact that equation (\ref{sddef}) is a valid steady state for $v \gg v_{ti}$ suggests that the increasing alpha particle density must be manifest where $v \sim v_{ti}$. This is the buildup of helium ash in local thermal equilibrium with the main ions. This cold, Maxwellian helium is known to get periodically ejected by the plasma during sawtooth crashes \citep{Nave2003}. 

In this work, we will not concern ourselves with the fate of the ash; we limit ourselves to the effects of and on the high-energy ($v \gg v_{ti}$) non-Maxwellian tail described by equation (\ref{sddef}). We will find that the transport properties of alpha particles depend strongly on energy, so we must be careful not to extend our conclusions to regimes in which our distribution is not valid, namely $v \sim v_{ti}$. Unless otherwise stated, the term ``alpha particles'' will refer to a non-Maxwellian species described by this slowing down distribution.

To solve the gyrokinetic equation, we also need $\nabla F_{S\alpha}$ as a function of velocity. To obtain this, we apply the chain rule to equation (\ref{sddef}), using the definition of $v_c$ in equation (\ref{vcdef}):
\begin{align} 
\frac{\nabla F_{S\alpha}}{F_{S\alpha}} =& \frac{1}{F_{S\alpha}} \frac{\partial F_{S\alpha}}{\partial n_\alpha} \nabla n_\alpha + \frac{1}{F_{S\alpha}} \frac{\partial F_{S\alpha}}{\partial v_c} \nabla v_c \label{gradFs} \\
=& \frac{\nabla n_\alpha}{n_\alpha} + \frac{1}{F_{S\alpha}} \frac{\partial F_{S\alpha}}{\partial v_c} \left( \frac{\partial v_c}{\partial T_e} \nabla T_e + \frac{\partial v_c}{\partial n_i} \nabla n_i + \frac{\partial v_c}{\partial n_e} \nabla n_e \right) \nonumber \\
\label{gradFs} =& \frac{\nabla n_\alpha}{n_\alpha} + \left[ \frac{v_\alpha^3}{v_c^3 + v_\alpha^3} \frac{1}{\ln \left(1 + v_\alpha^3/v_c^3 \right)} - \frac{v_c^3}{v_c^3 + v^3} \right] \times \nonumber \\
&\left( \frac{3}{2} \frac{\nabla T_e}{T_e} - \frac{\nabla n_e}{n_e} + \frac{\sum_i \, Z_i^2 \nabla n_i /m_i}{\sum_i\, Z_i^2 n_i /m_i}   \right). \nonumber
\end{align}
Choose a suitable radial coordinate $\rho$ such that $\nabla F_{0s} = \nabla \rho \partial F_{0s} / \partial \rho$ (in our simulations it is defined as the half-diameter of the flux surface at the height of the magnetic axis). It will be convenient to define the gradient length scale of $v_c$ thusly:
\begin{equation}\label{RLvc}
\frac{1}{L_{v_c}} \equiv \frac{\partial}{\partial \rho} \ln v_c  =  \frac{3}{2} \frac{1}{L_{T_e}} - \frac{1}{L_{n_e}} + \frac{\sum_i Z_i^2 n_i / m_i  L_{n_i} }{\sum_i Z_i^2 n_i / m_i }.
\end{equation}
In the flux-tube approximation, we assume that this quantity, and others defined similarly for $T_i$, $n_e$, etc., are constant across the simulation domain.

In this section, we have briefly discussed the gyrokinetic framework and how it is modified when species are allowed to be non-Maxwellian. In order to solve the gyrokinetic equation (\ref{gkeqn}) and associated field equation (\ref{qneutrality}), it will be necessary to keep $\partial F_{0s} / \partial E$ and $\nabla F_{0s}$ as more general functions of energy. This has been done with the \texttt{GS2} local flux-tube code, and in the following sections, we use this tool to test various assumptions made about alpha particles with self-consistent simulations. 

\subsection{Simulation setup}

\texttt{GS2} is a local flux-tube code that solves the gyrokinetic equation in field-following ``twist-and-shift'' coordinates \citep{Cowley1991}, and is capable of handling arbitrary equilibrium $F_{0s}$. The ``cyclone base case'' \citep{Dimits2000} is characterised by an $\hat{s}-\alpha$ geometry with an aspect ratio of $\epsilon = 0.4$ and a magnetic shear of $\hat{s} = 0.8$. At the radius $r/a = 0.45$, we take the field line pitch to be $q = 1.39$. The gradient scale lengths are: $R/L_{ni} = R/L_{ne} = 2.2$, and $R/L_{Ti} = R/L_{Te} = 6.9$. The ion temperature is $T_i = T_e = 10 \mathrm{keV}$. We use a grid of 32 points along the field line and for velocity space: 16 points in $v$ and 33 points in $\lambda$. When nonlinear simulations are run for this case, a perpendicular box size of $L_x \approx L_y = 63 \rho_i$ is used with a resolution of $N_x = N_y = 64$. The main ion species is deuterium, and electrons are assumed adiabatic: $\delta n_e / n_{e} \equiv \int \delta f_e d^3 \mathbf{v} / n_e = e \phi / T_e$. A linearized, conservative collision operator \citep[see][]{Abel2008,Barnes2009a} was used with $\nu_{ei} = 0.01 v_{ti}/a$.

Unless otherwise stated, the simulations in sections 3 and 4 have these parameters. The test case of section 5 is described therein.

\section{The trace-alphas approximation}

We inquire: at what concentration do alpha particles begin contributing to the turbulent dynamics? In any of the existing or planned fusion devices, the fraction $n_\alpha / n_e$ is expected to peak at most around 1\% \citep{Budny2002}. Considering that alpha particles have such high energy, it is not obvious whether or not they contribute to the electrostatic nonlinear dynamics of the plasma as these densities.  

When the density of a charged species is negligible, so is its contribution to the right hand side of the quasineutrality condition (\ref{qneutrality}). In the limit of $n_\alpha \rightarrow 0$ (holding $\delta n_\alpha / n_\alpha$ constant), $\phi$ no longer depends on the perturbation $h_\alpha$, in which case the gyrokinetic equation (\ref{gkeqn}) is linear in $h_\alpha$. We can therefore write the gyrokinetic equation (\ref{gkeqn}) as:
\begin{equation}\label{gkeqnlinop}
\mathcal{L} \left[ h_\alpha, \phi \right] =  - Z_\alpha e \frac{\partial F_{0\alpha}}{\partial E} \frac{\partial \left\langle \phi \right\rangle_{\mathbf{R}_s}}{\partial t} - \frac{c}{B} \mathbf{b} \times \nabla \left\langle \phi \right\rangle_{\mathbf{R}_s} \cdot \nabla F_{0\alpha},
\end{equation}
where $\mathcal{L}$ is the linear operator defined by the left hand side of equation (\ref{gkeqn}), and $\phi$ is treated as a given function of space and time, determined by the turbulent dynamics of the other, non-negligible species. Note that this does not imply that $h_\alpha$ is linear in $\phi$, nor that the usually-nonlinear $E \times B$-drift term in (\ref{gkeqn}) is ignored. Invert equation (\ref{gkeqnlinop}) to obtain $\delta f_\alpha$ and plug into equation (\ref{pfluxdef}). It follows that we can write the particle flux in the form \citep{Angioni2008}:
\begin{equation}\label{fluxcoeff}
\frac{R}{n_\alpha} \Gamma_\alpha = D \frac{R}{L_{n_\alpha}} + D_E \frac{R}{L_{T_e}} + V_p.
\end{equation}
From left to right, the terms are: particle diffusion, thermodiffusion, and the pinch flux (flux at zero gradient). The \emph{electron} temperature gradient appears here because that is the dominant dependence of the $v_c$ parameter when a single ion species is present with $Z_i = 1$ and $n_i \approx n_e$ (see equation (\ref{gradFs})).

\subsection{Linear theory}

\begin{figure}
\begin{center}
\includegraphics[width=0.5\textwidth]{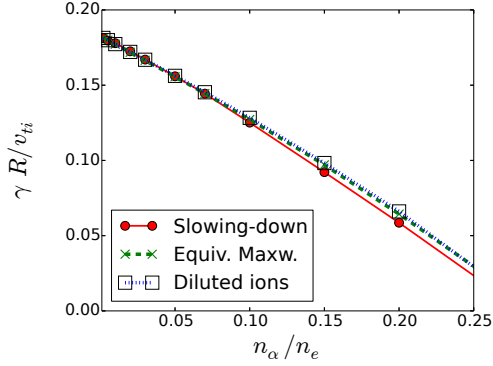}
\caption{ \label{linearf0comp} Comparison of linear growth rates for different models of alpha particles at a range of concentrations. Calculations were performed by running \texttt{GS2} for a single $k_y = 0.3$ mode of the cyclone base case \citep{Dimits2000} with $R/ L_{n\alpha} = R/L_{ni} = R/L_{ne} = 2.2$. Agreement between all three is within $1\%$ up to $n_\alpha / n_e \approx 0.05$, and still within $10\%$ up to an impossibly-large $n_\alpha / n_e \approx 0.15$. Note that the equivalent Maxwellian and diluted-ion models are nearly identical.}
\end{center}
\end{figure}

\begin{figure}
\begin{center}
\includegraphics[width=0.65\textwidth]{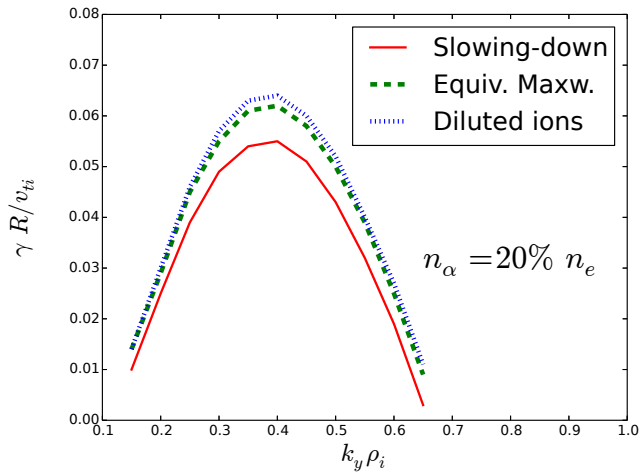}
\caption{ \label{kyspectrum} Growth rate spectrum of linear ITG growth rate at a 20\% alpha particle concentration. Same case as figure \ref{linearf0comp}. }
\end{center}
\end{figure}

A first estimate of how much of an effect alphas have on the plasma can be obtained by examining the linear growth rate of an unstable ion temperature gradient (ITG) mode. We examine the frequency and growth rate of the $k_y \rho_i = 0.4$ poloidal mode as alpha particles are introduced at ever-increasing density in figure \ref{linearf0comp}. The growth rate decreases with increasing alpha particle concentration, but only changes by about 5\% up to an alpha particle concentration of 2\%. Even at a concentration of $20\%$, we still see in figure \ref{kyspectrum} that there is no qualitative and little quantitative difference in the poloidal spectrum. We find that the effect of a small population of alpha particles is negligible, at least linearly.

As $n_\alpha / n_e$ increases, the relative fraction of main ions (whose temperature gradient drives the instability) must decrease to compensate and maintain equilibrium quasineutrality, resulting in a dilution effect \citep[see][]{Tardini2007, Holland2012a}. It could be argued whether this effect alone is responsible for the change in growth rate shown in figures \ref{linearf0comp} and \ref{kyspectrum}. Therefore, what is also shown (labelled ``diluted ions'') is the case where alphas do not contribute to the field at all, even at significant density. Indeed, it takes very high concentrations of alpha particles ($\gtrapprox 10 \%$) to distinguish between the different models (see section 4 for an explanation of the ``equivalent Maxwellian'' model), and no model at all. This suggests that, even beyond realistic reactor densities, the primary effect of alpha particles is only to dilute the ITG-driving ions, introducing no particularly interesting electrostatic effects of their own. 

\subsection{Nonlinear simulations}

\begin{figure}
\begin{center}
\includegraphics[width=0.65\textwidth]{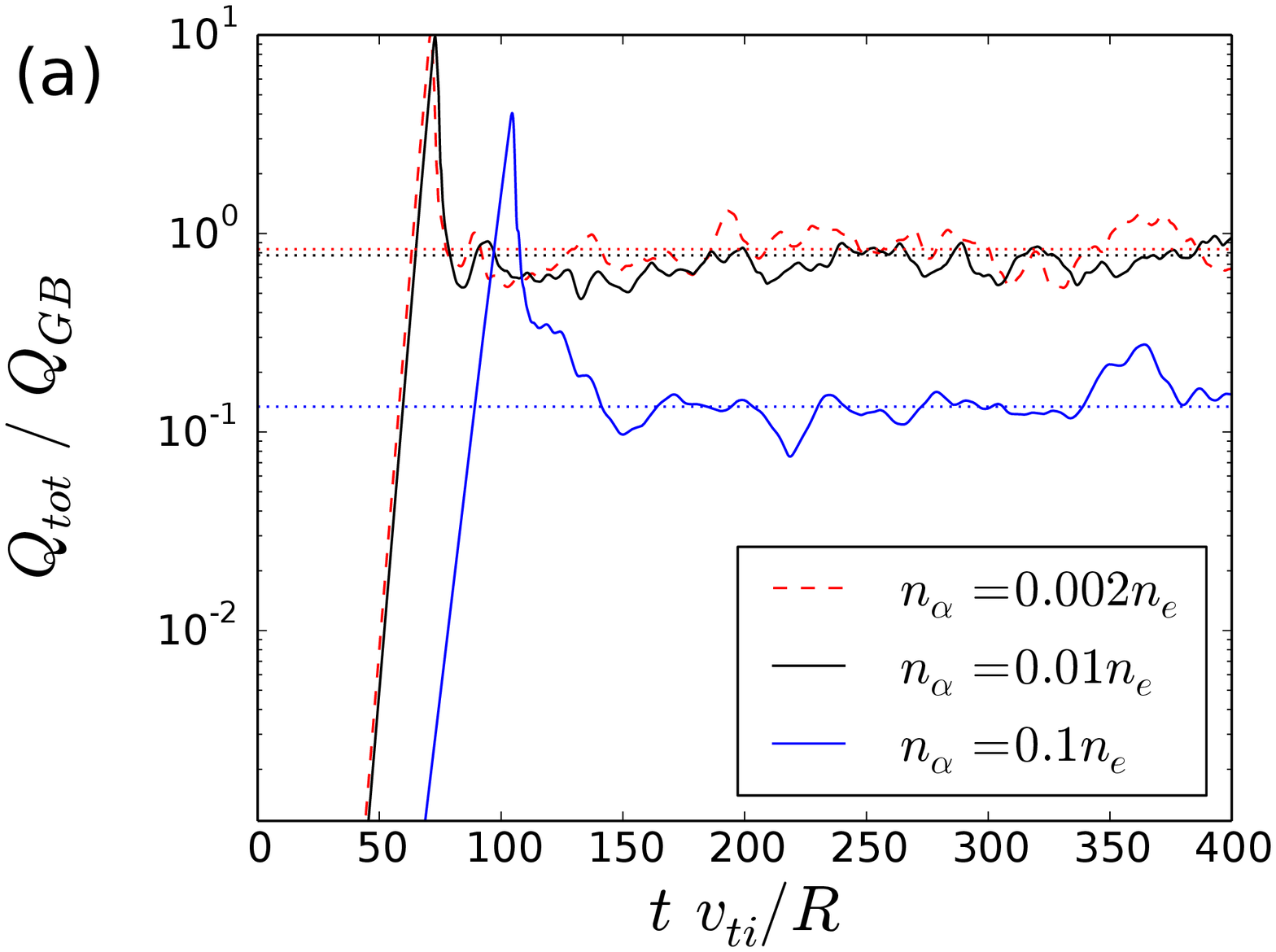}
\includegraphics[width=0.65\textwidth]{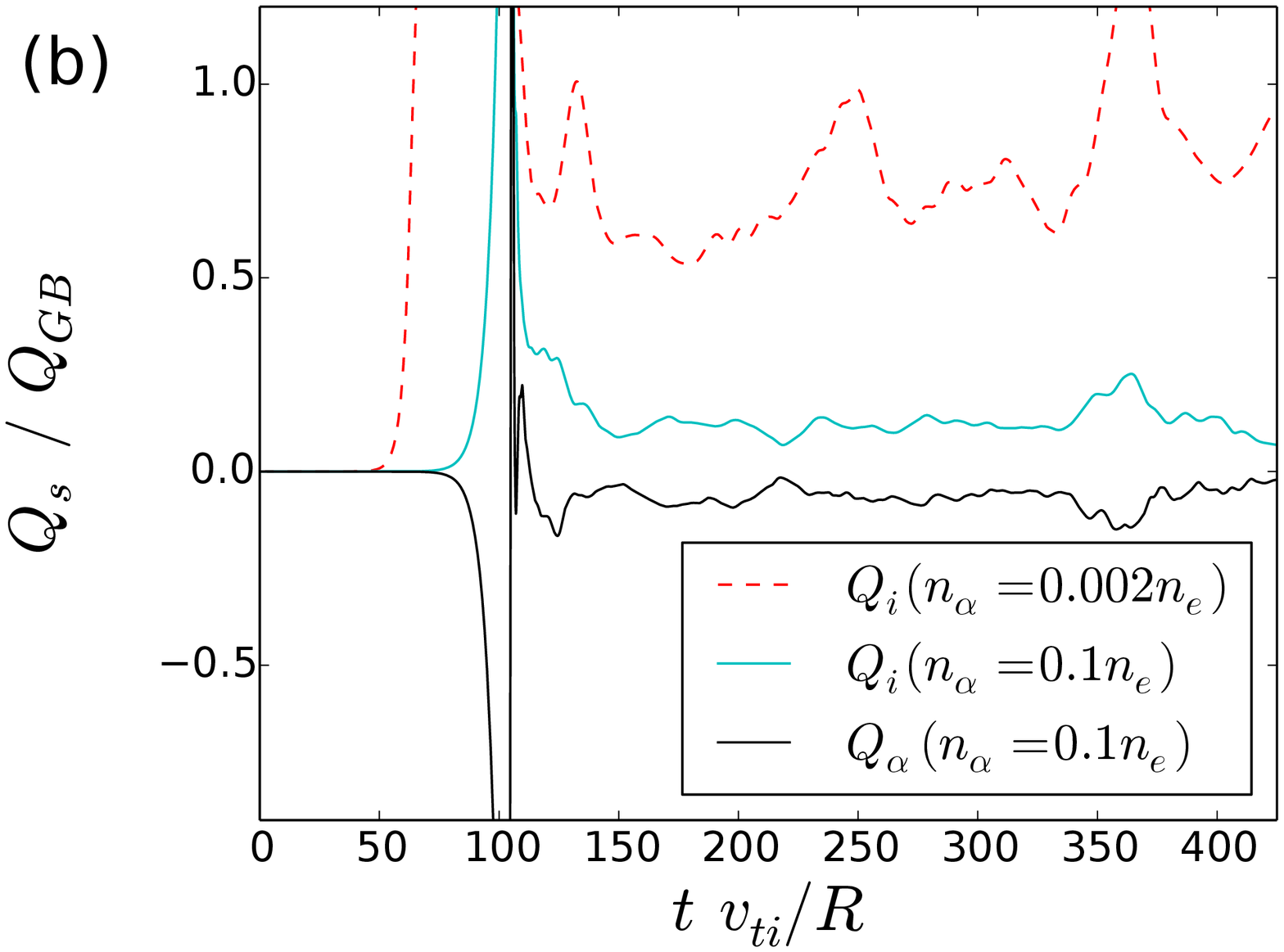}
\caption{\label{hfluxvst} Time evolution of the turbulent heat flux. The dotted horizontal lines are the time-averaged heat fluxes for the different concentrations of alpha particles. $Q_{GB} \equiv n_e v_{ti} T_i \rho^{*2}$. (\textit{a}) shows the total heat flux for different alpha particle concentrations, and (\textit{b}) shows the breakdown by species at $n_\alpha / n_e = 0.1$, compared to the ion heat flux at negligible alpha density. }
\end{center}
\end{figure}

We then proceed to demonstrate that these conclusions continue to hold in turbulence. We turn on the nonlinear term in equation (\ref{gkeqn}) and examine the evolution of fluxes to an approximate steady-state. The time evolution to saturation of the total heat flux is shown in figure \ref{hfluxvst}. In this case, the decrease in outward total heat flux is due to the combined effect of: 1) alpha particles carrying some heat inward; and 2) reducing the ITG drive by the main ions. An inward heat and particle flux for the alpha particles is seen because there is an inward flux of alphas due to the second two terms in equation \ref{fluxcoeff}, but the alpha density gradient is not strong enough in this case (with $R/L_{n\alpha} = R/L_{ni}$) for the diffusion term to dominate and drive the alpha particles outward.

\begin{figure}
\begin{center}
\includegraphics[width=0.65\textwidth]{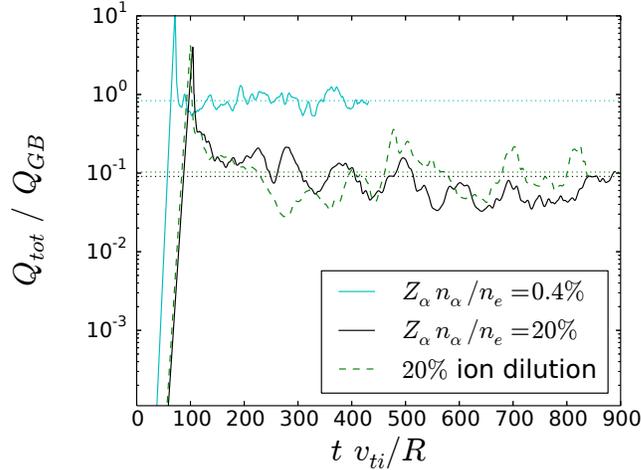}
\caption{\label{hfluxdilute} Time evolution of the turbulent heat flux, comparing the case of: a small alpha population (solid cyan), a large alpha population (solid black), and a case where the presence of alphas is ``simulated'' only be diluting the ion density (dashed green)}
\end{center}
\end{figure}

Even at high concentrations of alpha particles ($\sim 10\%$), the effect on the turbulence is indistinguishable from that of mere dilution of the main ions, consistent with linear theory. This is demonstrated in figure \ref{hfluxdilute}, which shows only a $10\%$ difference between heat fluxes between the case with alpha particles and that without, wherein the latter of which only the ion dilution effect is taken into account.

\begin{figure}
\begin{center}
\includegraphics[width=0.5\textwidth]{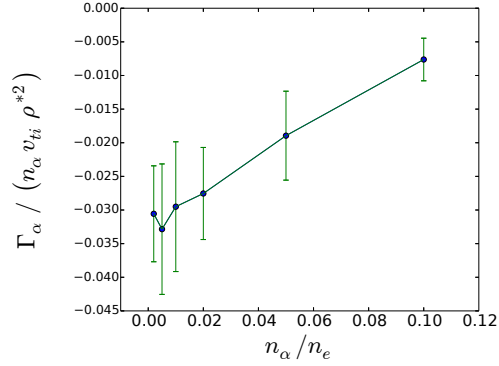}
\caption{\label{pfluxvsna} Steady-state turbulent flux of alpha particles as a function of alpha particle concentration. Units are gyrobohm normalised by the alpha particle density. ``Error bars'' indicate the standard deviation of the departure of fluctuations from the time average, and is intended to put into context the variations of flux at low concentration.}
\end{center}
\end{figure}

To see the effect this has on alpha transport, let us also compare the alpha particle flux. If alphas have little or no effect on the turbulence, we would then expect $\Gamma_{\alpha} / n_\alpha$ to be constant as the concentration changes. The time-averaged value of $\Gamma_\alpha / n_\alpha$ compared to alpha particle concentration is shown in figure \ref{pfluxvsna}. It is clear that no significant change occurs below a concentration of about 5\%.

\begin{figure}
\begin{center}
\includegraphics[width=0.65\textwidth]{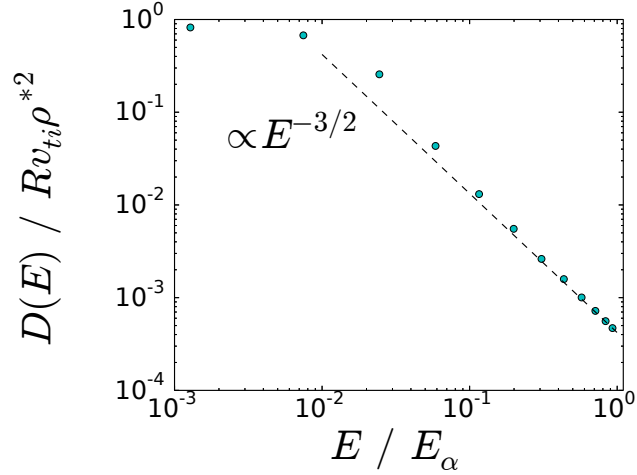}
\caption{\label{diffCoeffVSe} Diffusion coefficient of trace alpha particles ($n_\alpha = 0.002 n_e$) as a function of energy. }
\end{center}
\end{figure}

As mentioned previously, by assuming energetic ions are of negligible density, it can be shown \citep[see][]{Hauff2009,Pueschel2012} that for almost all pitch angles (the dependence on which is not covered in this work since we are focusing on isotropic alpha particles), the diffusion coefficient scales like $E^{-3/2}$. To make sense of this quantity, consider the energy-dependent analogue of equation (\ref{fluxcoeff}) consistent with the energy-dependent flux of equation (\ref{fluxedef}):
\begin{equation}\label{fluxecoeff}
\frac{R}{F_{0\alpha}(E)} \Gamma_\alpha(E) = D(E) \frac{R}{L_{n_\alpha}} + D_E(E) \frac{R}{L_{T_e}} + V_p(E).
\end{equation}
Note that, with this definition, $D(E)$ has the same units as its energy-integrated counterpart. By performing several nonlinear runs with a range of density gradients, a linear fit of $\Gamma_\alpha(E)$ versus $R/L_{n_\alpha}$ was performed, the slope of which is proportional to the diffusion coefficient $D(E)$. The results are plotted in figure \ref{diffCoeffVSe}, with a scaling and approximate magnitude consistent with \citet{Hauff2009,Pueschel2012}.

The conclusion of this section is to confirm that in the presence of electrostatic turbulence, an energetic species has little effect up to a concentration of at least $2\%$. However, even beyond such a density, they do not have much of a direct effect on the turbulence. Instead, their effect is simply to dilute the main ions, decreasing the ITG drive. This dilution effect is the dominant influence of fast ions up to at least a concentration of 10\%.

\section{The equivalent-Maxwellian approximation}

Even though a concentration of fast ions appears to have little effect on electrostatic turbulence, the response of alpha particles to that turbulence depends quite explicitly on the equilibrium distribution function, and especially its radial gradient. 

To take advantage of existing tools to solve the gyrokinetic equation for Maxwellian equilibria, it is naturally suggested that, instead of representing the alpha particles with a slowing-down distribution (e.g. equation (\ref{sddef})), one could instead define a Maxwellian that has the same temperature, and that this may provide satisfactory results. This method of modelling alpha particles has been widely used in gyrokinetic studies of alpha particles \citep[see, e.g.][]{Estrada-Mila2006,Angioni2009, Nishimura2009, Albergante2009, Zhang2010,Pueschel2012, Citrin2013a, Mishchenko2014}.

\subsection{Definitions}

The zeroth and second moment of the slowing-down distribution, equation (\ref{sddef}), can be evaluated analytically. We use these to define an effective temperature $T_{\text{eff}} = T_{\text{eff}}\left( v_c / v_\alpha \right)$ such that:
\begin{equation}\label{Teffdef}
\frac{3}{2} n_\alpha T_{\text{eff}} = \int F_{M \alpha} m_\alpha v^2 \, d^3 \mathbf{v} = \int F_{S \alpha} m_\alpha v^2 \, d^3 \mathbf{v},
\end{equation}
so \citep{Estrada-Mila2006}:
\begin{align}
T_{\text{eff}} =& \frac{ m_\alpha  v_c^2}{2  \ln\left( 1 + v_\alpha^3/v_c^3 \right)}\times \label{Teff} \\
& \left[ \frac{v_\alpha^2}{v_c^2} - \frac{1}{3} \ln \left( \frac{ v_\alpha^2 - v_\alpha v_c + v_c^2}{\left( v_\alpha + v_c \right)^2} \right) - \frac{1}{2 \sqrt{3}} \tan^{-1} \left( \frac{2 v_\alpha - v_c}{\sqrt{3} v_c} \right) - \frac{\pi}{3 \sqrt{3}} \right] . \nonumber
\end{align}

However, the gradient of the equivalent Maxwellian also appears in the gyrokinetic equation (\ref{gkeqn}), so we need a way of calculating it. Fortunately, $v_c$ (hence $T_\text{eff}$) is a known function of ion and electron parameters, so we can find the effective Maxwellian temperature gradient scale length by using the chain rule, in a manner analogous to equation (\ref{gradFs}):
\begin{equation}\label{gradFMeff}
\frac{\nabla F_{M \alpha}}{F_{M \alpha}} = \frac{\nabla n_\alpha}{n_\alpha} + \frac{1}{T_{\text{eff}}} \frac{\mathrm{d} T_{\text{eff}}}{\mathrm{d} v_c}  \left( \frac{\partial v_c}{\partial T_e} \nabla T_e + \frac{\partial v_c}{\partial n_i} \nabla n_i + \frac{\partial v_c}{\partial n_e} \nabla n_e \right)  \left( \frac{E}{T_\text{eff}} - \frac{3}{2} \right).
\end{equation}
We take the derivative of equation (\ref{Teff}) and write down an expression for $R / L_{T_\text{eff}}$ as a function of $x \equiv v_c / v_\alpha$:
\begin{align}\label{RLTeff}
\frac{R}{L_{T_{\text{eff}}}} = \frac{R}{L_{v_c}} \frac{1}{\ln \left( 1 + x^{-3} \right)} &\left[  \frac{1}{1 + x^3} - \frac{x^2}{3} \left(\frac{E_\alpha}{T_{\text{eff}}}\right)  \frac{2 \pi}{3 \sqrt{3}} + \frac{2}{\sqrt{3}} \tan^{-1} \left( \frac{2- x}{\sqrt{3} x} \right) \right. \nonumber \\
&+ \left. \frac{1}{3} \ln \left( \frac{ x^2 - x + 1}{\left( 1 + x \right)^2} \right) - \frac{x}{\left(1+x\right) \left(x^2 - x + 1 \right)}  \right]
\end{align}
where $R / L_{v_c}$ is given by equation (\ref{RLvc}).

\subsection{Linear theory}

Proceeding in a manner analogous to section 3, we analyze the linear mode that results from using a concentration of alpha particles using the equivalent Maxwellian versus the slowing-down distribution. Consistent with \citet{Estrada-Mila2006} (which used a different test case), we find that the growth rates for the slowing-down and Maxwellian distributions in the cyclone base case follow each other very closely up to relatively high concentration (see figure \ref{linearf0comp}). This is unsurprising given the conclusion of the previous section: that a modest concentration of alpha particle plays no electrostatic role except dilution.

We proceed to ask the inverse question: how do alpha particles respond to a given linearly unstable eigenfunction, and how does the equilibrium distribution function used affect the result? We can use quasilinear theory to estimate the fluxes with the same method as \citet{Angioni2008}. That is, for each set of parameters, we choose a single unstable mode and calculate the alpha particle flux (equation \ref{pfluxdef}) as a function of time. Because it is exponentially growing, we must normalize it to a quantity growing at the same rate, such as the flux of ash (a helium species at the same temperature as the ions). This only works because in both cases, the density is taken to be trace, otherwise there would be a small but catastrophic difference in growth rates. This ratio of alpha flux to ash flux in response to the linear eigenfunction is what we calculate.

\begin{figure}
\begin{center}
\includegraphics[width=0.45\textwidth]{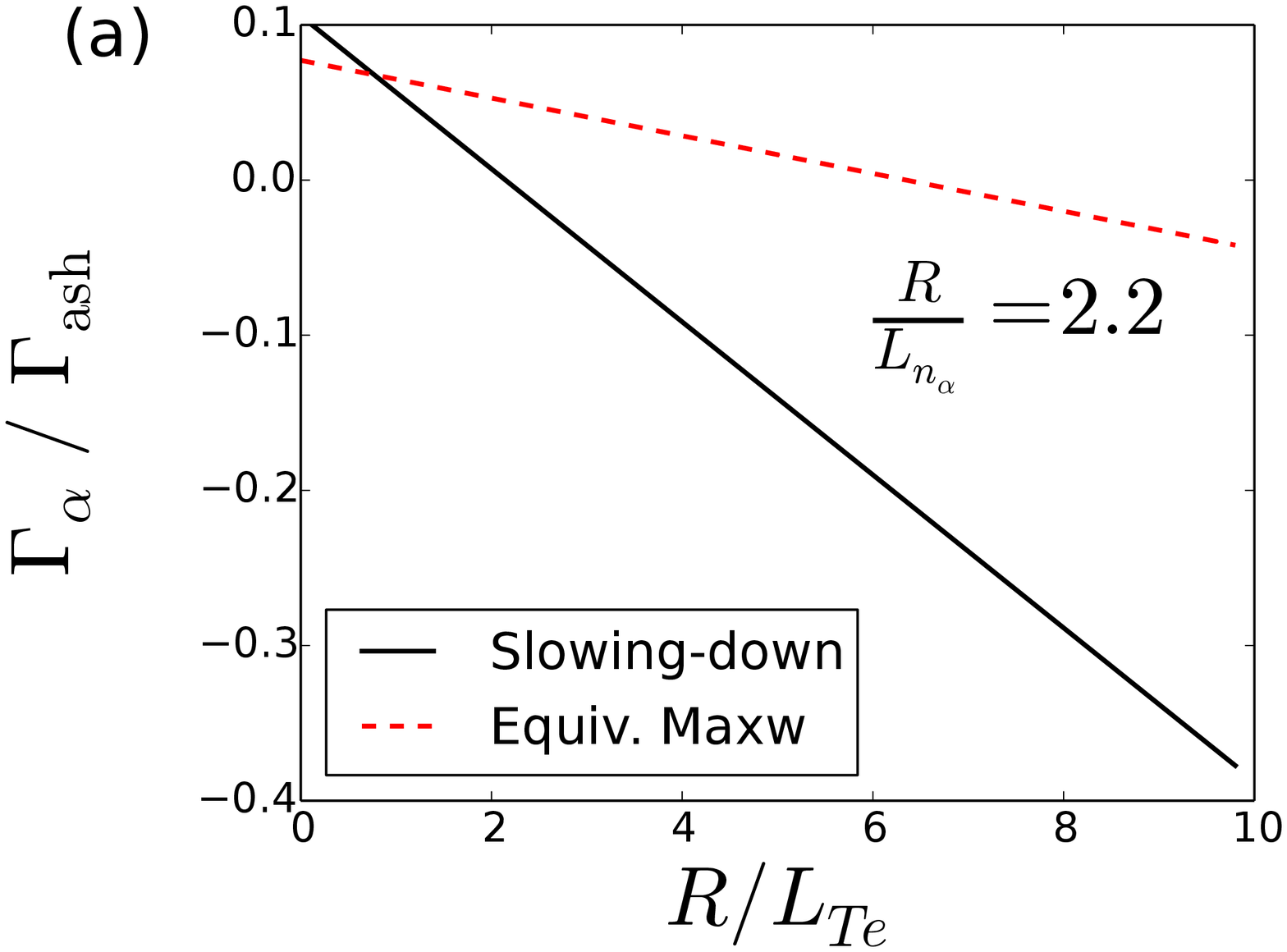}
\includegraphics[width=0.45\textwidth]{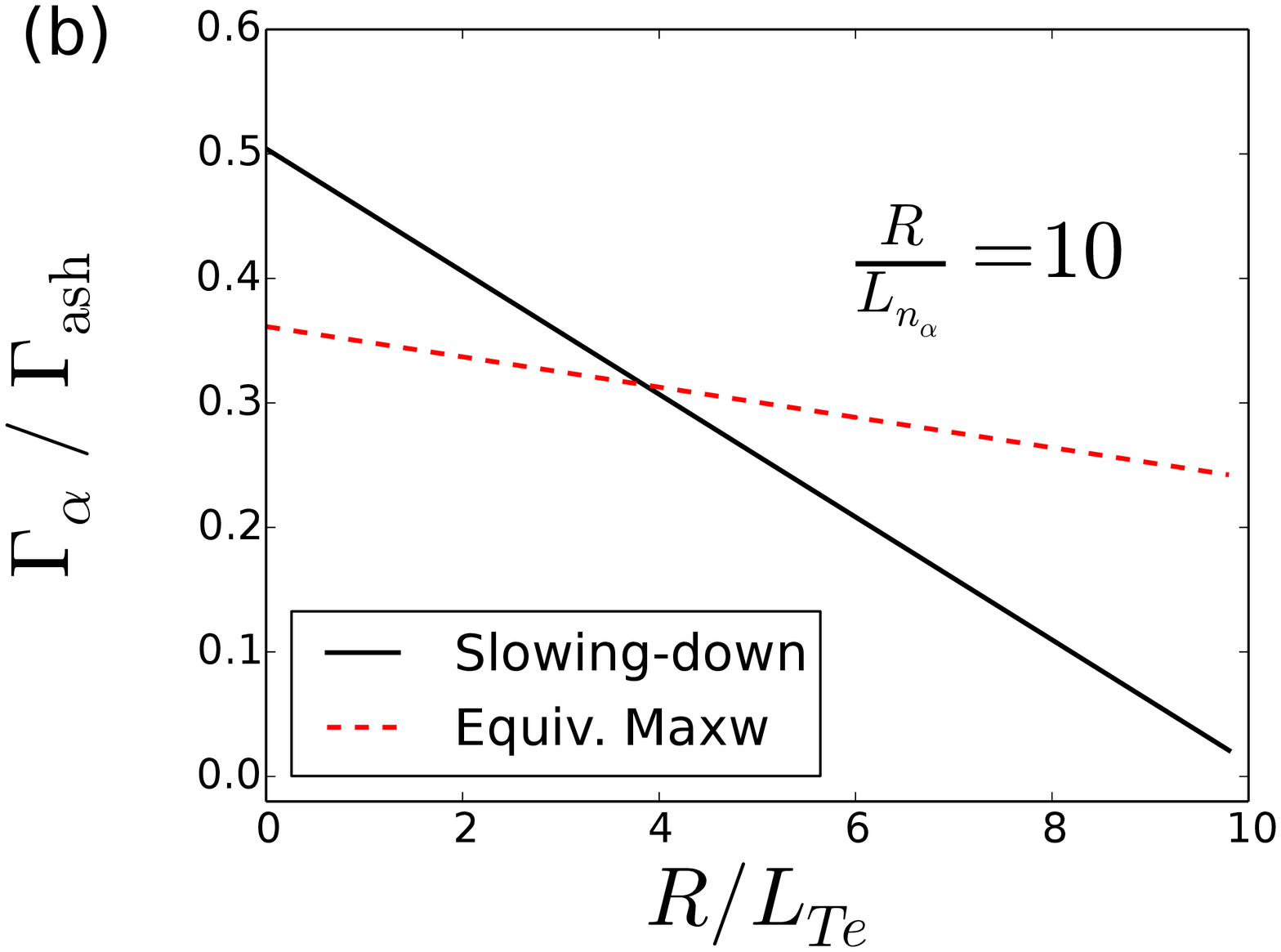}
\caption{\label{qlinscan} Quasilinear alpha particle flux determined by finding linear fits for the coefficients in equation (\ref{fluxcoeff}). Showing the dependence on the electron temperature gradient for (\textit{a}) $R/L_{n_\alpha} = R/L_{n_i} =  2.2$, and (\textit{b}) $R/L_{n_\alpha} = 10$.}
\end{center}
\end{figure}

Consider again the fact that, in the trace limit, the gyrokinetic equation (\ref{gkeqnlinop}) is linear in the gradients. Then, equation (\ref{fluxcoeff}) holds, and the particle flux is easily found after finding the coefficients $D$, $D_E$, and $V_p$. After fitting these coefficients to a series of linear simulations for the cyclone case, we plot the dependence of particle flux on the dominant parameters ($R / L_{n_\alpha}$ and $R / L_{T_e}$) in figure \ref{qlinscan}. From inspection, one can see that, depending on the problem parameters, one can achieve anything from very good to very poor agreement between the slowing-down distribution and the equivalent Maxwellian. 

\subsection{Explanation of discrepancy}

\begin{figure}
\begin{center}
\includegraphics[width=0.5\textwidth]{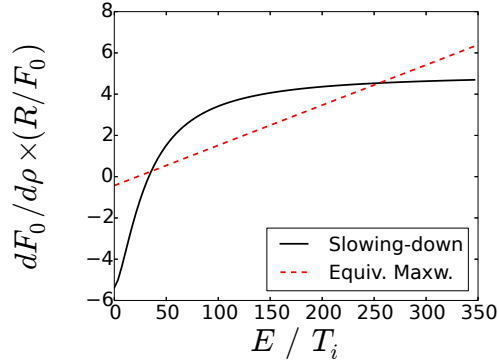}
\caption{\label{gradF0vsE} Comparing the energy dependence of the radial spatial derivative of $F_{0\alpha}$ for the two models of alpha particle distributions. In both cases, the gradient is found from the density gradient of alpha particles, and the gradient of $v_c$, which can be found as a function of the equilibrium parameters of other species, particularly $T_e$.  }
\end{center}
\end{figure}

To explain this disagreement, consider the $\nabla F_{0\alpha}$ term in the gyrokinetic equation (\ref{gkeqn}). For a Maxwellian distribution, $\nabla F_{M\alpha} / F_{M\alpha}$ is linear in energy (see equation (\ref{gradFMeff})), but the gradient of the slowing-down distribution has a different energy dependence (equation (\ref{gradFs})). From figure \ref{gradF0vsE}, we see that when $E \sim T_i$ (near which the interaction with the ion-scale turbulence is expected to be the strongest), the gradient of $F_{0\alpha}$ is off by over an order of magnitude. This stark difference in the right hand side of equation (\ref{gkeqn}) ultimately carries through to the particle flux, resulting in the discrepancies in figures \ref{qlinscan} and \ref{equivmaxwhflux}(\textit{b}). 

We conclude that the equivalent Maxwellian approximation is wrong precisely because it fails to capture the energy dependence of $\nabla F_{0\alpha}$, and at least sometimes strongly disagrees at the most relevant energies. In the diffusive limit, where the gradient of $F_{0\alpha}$ is dominated by $\nabla n_\alpha$, this energy dependence is not important, and one would expect the equivalent Maxwellian to predict at least the correct order of magnitude. Even so, if one wishes to find the gradient $R/L_{n\alpha}$ that eliminates the alpha particle flux, the balance with $D_E$ (which \emph{is} sensitive to the energy dependence of $\nabla F_{S\alpha}$) and $V_p$ in equations (\ref{fluxcoeff}) and (\ref{fluxecoeff}) is necessary.

\subsection{Nonlinear simulations}

\begin{figure}
\begin{center}
\includegraphics[width=0.65\textwidth]{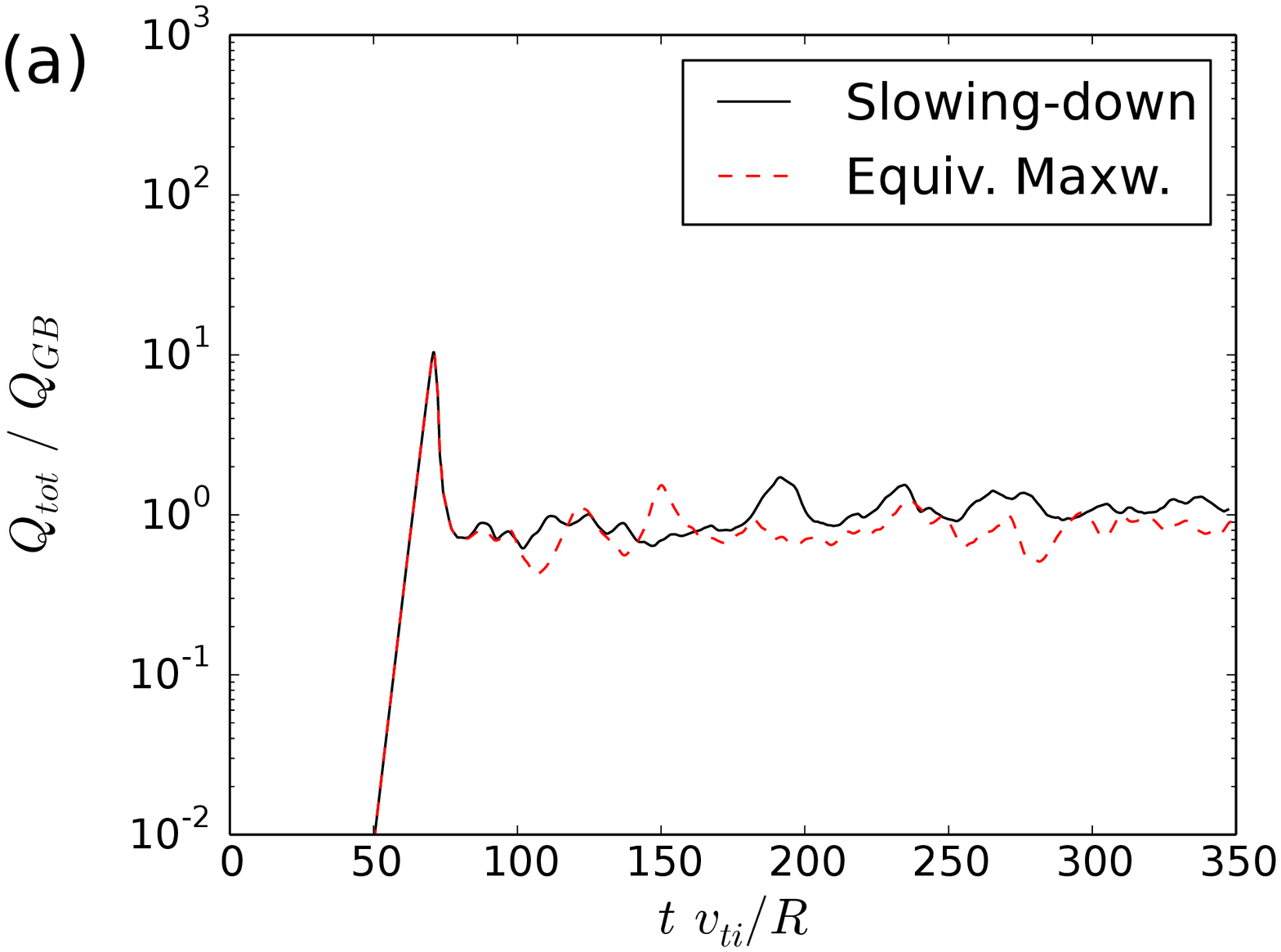}
\includegraphics[width=0.65\textwidth]{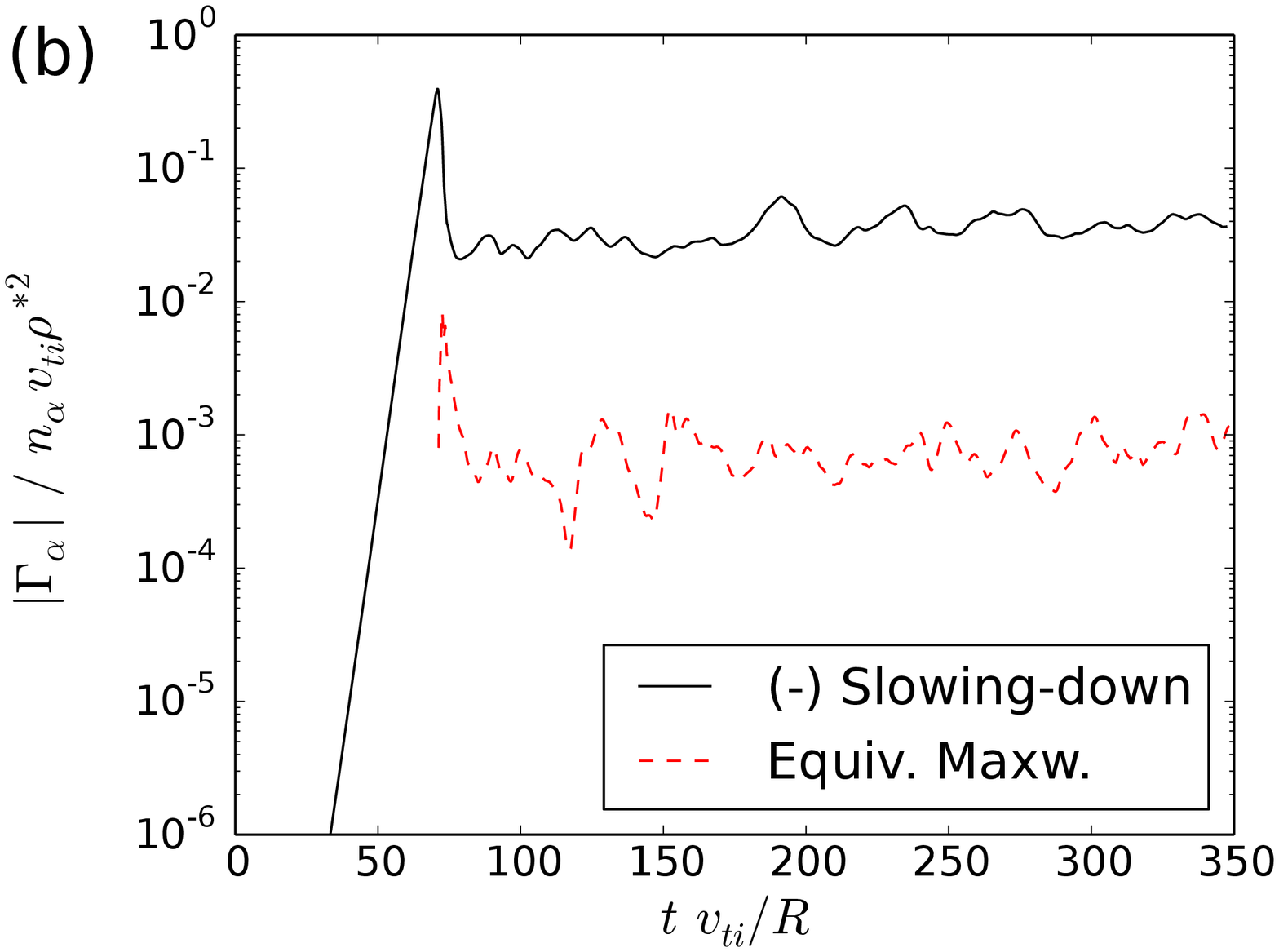}
\caption{\label{equivmaxwhflux} Time evolution of (\textit{a}) total heat flux, and (\textit{b}) alpha particle flux. Comparing two models for the alphas particles: the slowing down distribution (solid black), and the equivalent Maxwellian (dashed red). In both cases, the alpha particle concentration is 0.1\%. The negative particle flux for the slowing-down distribution is shown, since the signs do not agree.}
\end{center}
\end{figure}

One can see in figure \ref{qlinscan}(a) that the ``fully mixed'' cyclone base case ($R / L_{n_\alpha} = 2.2$, $R/L_{T_e} = 6.9$) is a particularly poorly performing case for the equivalent Maxwellian. Using these parameters, figure \ref{equivmaxwhflux}(a) compares the total heat flux for the two distributions. Since they are both well below the threshold to be considered ``trace'', there is little statistical difference in the total heat flux between these two, as would be expected. However, the turbulent fluxes shown in figure \ref{equivmaxwhflux}(b) demonstrate that the equivalent Maxwellian gets the wrong direction of the alpha particle flux flux and is off by more than an order of magnitude. 

While we do not claim that this strong of a disagreement will be seen in all relevant cases, the observation that: a) such an agreement is so sensitive to the parameters of the problem; b) a drastic difference is found for such a common test case as cyclone; and c) that such an agreement, when it does exist by coincidence, has no physical basis, should be enough to convince the reader that any results for alpha particle flux obtained by using an equivalent-Maxwellian ought to be treated with skepticism. Any disagreement in the fluxes is especially troublesome if one is performing a critical-gradient analysis to determine the alpha particle profile. From inspection of figure \ref{qlinscan}, one can observe very different critical gradients (the gradient for which $\Gamma_\alpha \rightarrow 0$) between the two distribution functions.

\subsection{Correcting the equivalent Maxwellian}

We can take advantage of the trace approximation to decompose the energy dependence of the transport coefficients given in equation (\ref{fluxecoeff}). Doing so will allow one to obtain the fluxes that one would get from a simulation with the slowing-down distribution, provided the coefficients $D$, $D_E$ and $V_p$ are known from a series of equivalent Maxwellian simulations.

Consider again the linearity of (\ref{gkeqnlinop}). Decompose the right hand side into terms with the known velocity dependences of $\partial F_0 / \partial E$ and $\nabla F_0$ factored out. When using the slowing-down distribution, we can write the gyrokinetic equation as:
\begin{equation}\label{gkdecomps}
\frac{1}{F_{S\alpha}} \mathcal{L}\left[h_\alpha \right] = a_0 M_0^{(S)} + a_1 M_1^{(S)} \frac{R}{L_{n_\alpha}} + a_2 M_2^{(S)} \frac{R}{L_{v_c}}.
\end{equation}
Analogously for the equivalent Maxwellian:
\begin{equation}\label{gkdecompm}
\frac{1}{F_{M\alpha}} \mathcal{L}\left[h_\alpha \right] = a_0 M_0^{(M)} + a_1 M_1^{(M)} \frac{R}{L_{n_\alpha}} + a_2 M_2^{(M)} \frac{R}{L_{v_c}}.
\end{equation}
The following quantities are defined:
\begin{align}
M^{(S)}_0 &= - E_\alpha \frac{\partial }{\partial E} \ln F_{S\alpha} = \frac{3}{2} \frac{ v^2 v_\alpha}{v_c^3 + v^3} \\
M^{(M)}_0 &= - E_\alpha \frac{\partial }{\partial E} \ln F_{M\alpha} = \frac{E_\alpha}{T_\mathrm{eff}} \\
M^{(S)}_1 &= M^{(M)}_1 = 1 \\
M^{(S)}_2 &= v_c \frac{\partial }{\partial v_c} \ln F_{S\alpha} = \frac{3 v_\alpha^3}{v_c^3 + v_\alpha^3} \frac{1}{\ln\left( 1+ v_\alpha^3/v_c^3 \right)} - \frac{3 v_c^3}{v_c^3 + v^3}\\
M^{(M)}_2 &= v_c \frac{\partial T_\mathrm{eff}}{\partial v_c} \frac{\partial}{\partial T_\mathrm{eff}} \ln F_{M\alpha} = \left( \frac{E}{T_{\mathrm{eff}}} - \frac{3}{2} \right) \frac{L_{v_c}}{L_{T_\mathrm{eff}}},
\end{align}
where $L_{v_c}$ and $L_{T_\mathrm{eff}}$ are given by equations (\ref{RLvc}) and (\ref{RLTeff}) respectively. These factors together contain the only dependence on $F_0$ that appear in the gyrokinetic equation. The other factors $a_0$, $a_1$, and $a_2$ are allowed to depend on velocity, but not through $F_0$. Therefore, these factors are the same in both equations (\ref{gkdecomps}) and (\ref{gkdecompm}), and the dependence on the equilibrium distribution is entirely contained in the \emph{a priori}-known functions $M_0$, $M_1$, and $M_2$.

\begin{figure}
\begin{center}
\includegraphics[width=0.7\textwidth]{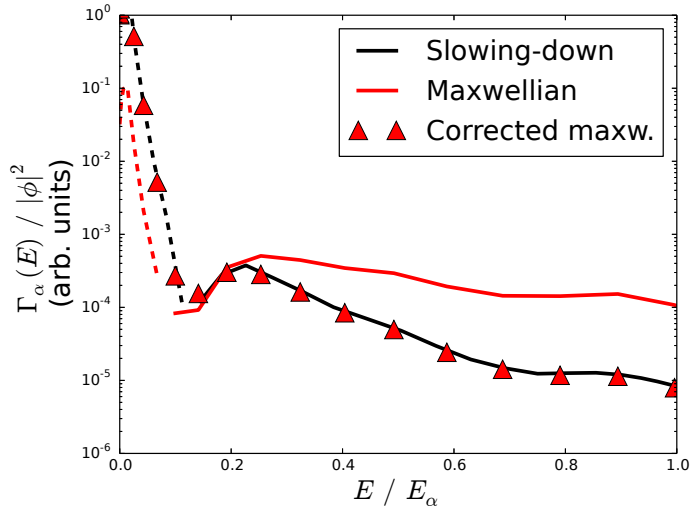}
\end{center}
\caption{ \label{fixmaxw} Quasilinear radial flux for the $k_y \rho_i = 0.3$ mode normalized to the total amplitude of $\phi$. Triangles represent adjustments made directly to the equivalent Maxwellian via equations (\ref{Dconvert})-(\ref{Vpconvert}). Dashed lines represent negative values.  }
\end{figure}

Suppose we know, from a simulation campaign using the equivalent Maxwellian approximation, the energy-dependent diffusion coefficients $D^{(M)}(E)$, $D_E^{(M)}(E)$, and $V_p^{(M)}(E)$. We can find the corresponding turbulent transport coefficients $D^{(S)}(E)$, $D_E^{(S)}(E)$, $V_p^{(S)}(E)$, and hence the radial flux $\Gamma_\alpha(E)$ for the slowing-down distribution, even if a gyrokinetic simulation with $F_{S\alpha}$ was never run. To convert between the two:
\begin{align}
D^{(S)}(E) &= D^{(M)}(E) \label{Dconvert} \\
D_E^{(S)}(E) &= \frac{M_2^{(S)}}{M_2^{(M)}} D_E^{(M)}(E) \label{DEconvert} \\
V_p^{(S)}(E) &= \frac{M_0^{(S)}}{M_0^{(M)}} V_p^{(M)}(E)  \label{Vpconvert} ,
\end{align}
and apply equation (\ref{fluxecoeff}). For the case whose nonlinear particle flux is shown in figure \ref{equivmaxwhflux}, these relationships were applied to the quasilinear flux of the fastest-growing mode. Figure (\ref{fixmaxw}) shows the $\Gamma_\alpha(E)$ that results when the equivalent Maxwellian is corrected. 

Note that the energy-dependent diffusion coefficient is identical between the two distributions. This is by construction in the way we have defined it in equation (\ref{fluxecoeff}) (following \citet{Angioni2008,Hauff2009,Pueschel2012}). This also explains why there is better agreement in $\Gamma_\alpha(E)$ at higher gradients: the diffusion term tends to dominate. However, to obtain a reasonable estimate for the integrated particle flux in this regime, one must normalize properly by $F_{S\alpha}$. Again, we caution against taking this diffusive approximation too far, for example, in performing a critical gradient analysis, where the other terms in equation (\ref{fluxecoeff}) do indeed become important. However, the use of the conversions (\ref{Dconvert})-(\ref{Vpconvert}) should be adequate even for this purpose so long as alpha particles remain trace.

\section{Confinement of alpha particles in ITG turbulence}

Implicit in the use of the slowing-down distribution (equation (\ref{sddef})) is the assumption that alpha particles are well-confined in the sense that all collisional slowing-down happens on approximately the same flux surface: that the particle transport time is long compared to the slowing-down time. In this section, we will analyze this assumption and its associated subsidiary expansion introduced in section 2 using the results from a nonlinear local ITER simulation. Said analysis will be \emph{a posteriori}: assuming a classical slowing-down velocity distribution for fast alpha particles, how likely is it that it remains so when taking into account turbulent transport? 

\subsection{Test case}

\begin{figure}
\begin{center}
\includegraphics[width=0.4\textwidth]{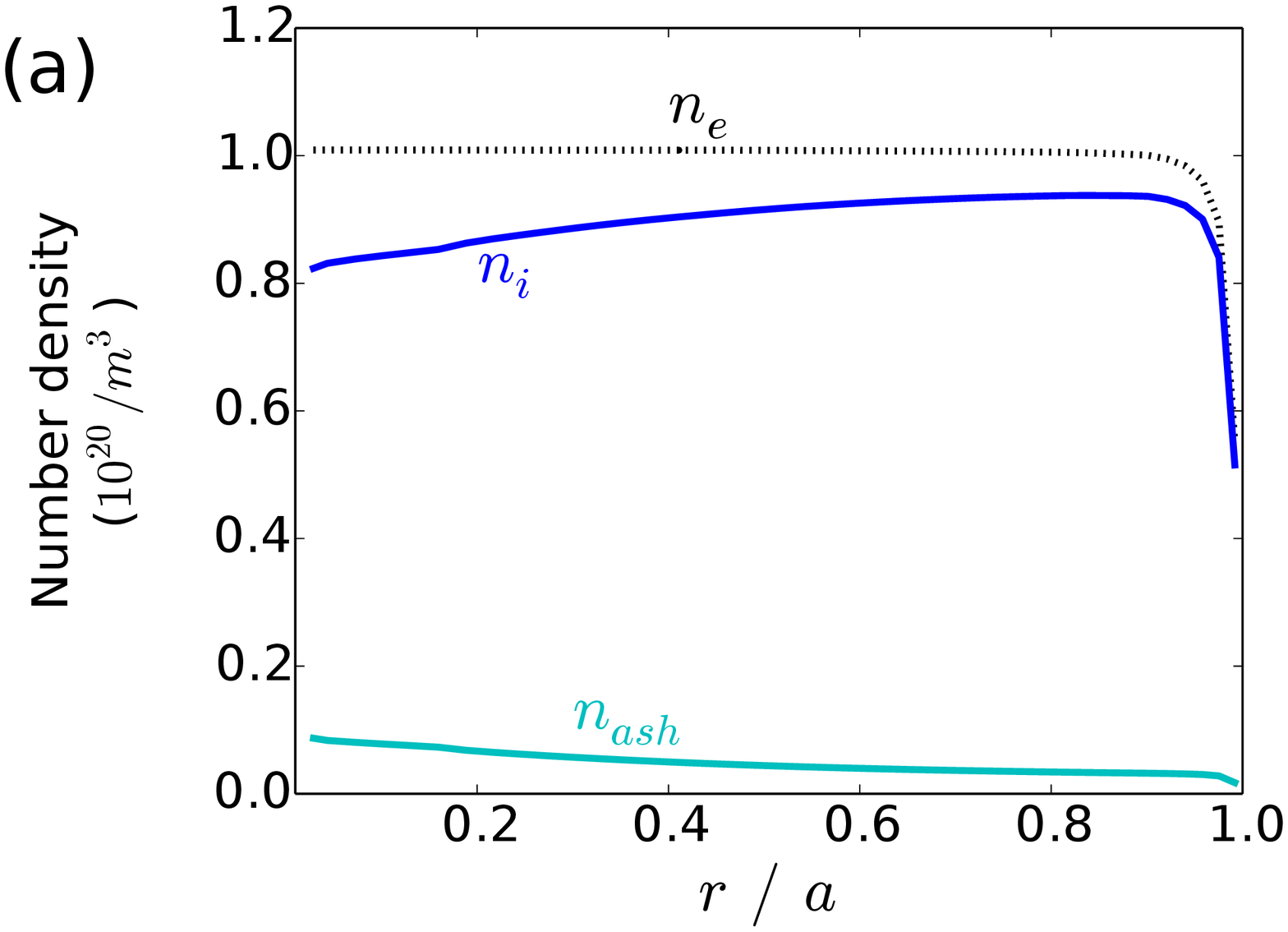}
\includegraphics[width=0.4\textwidth]{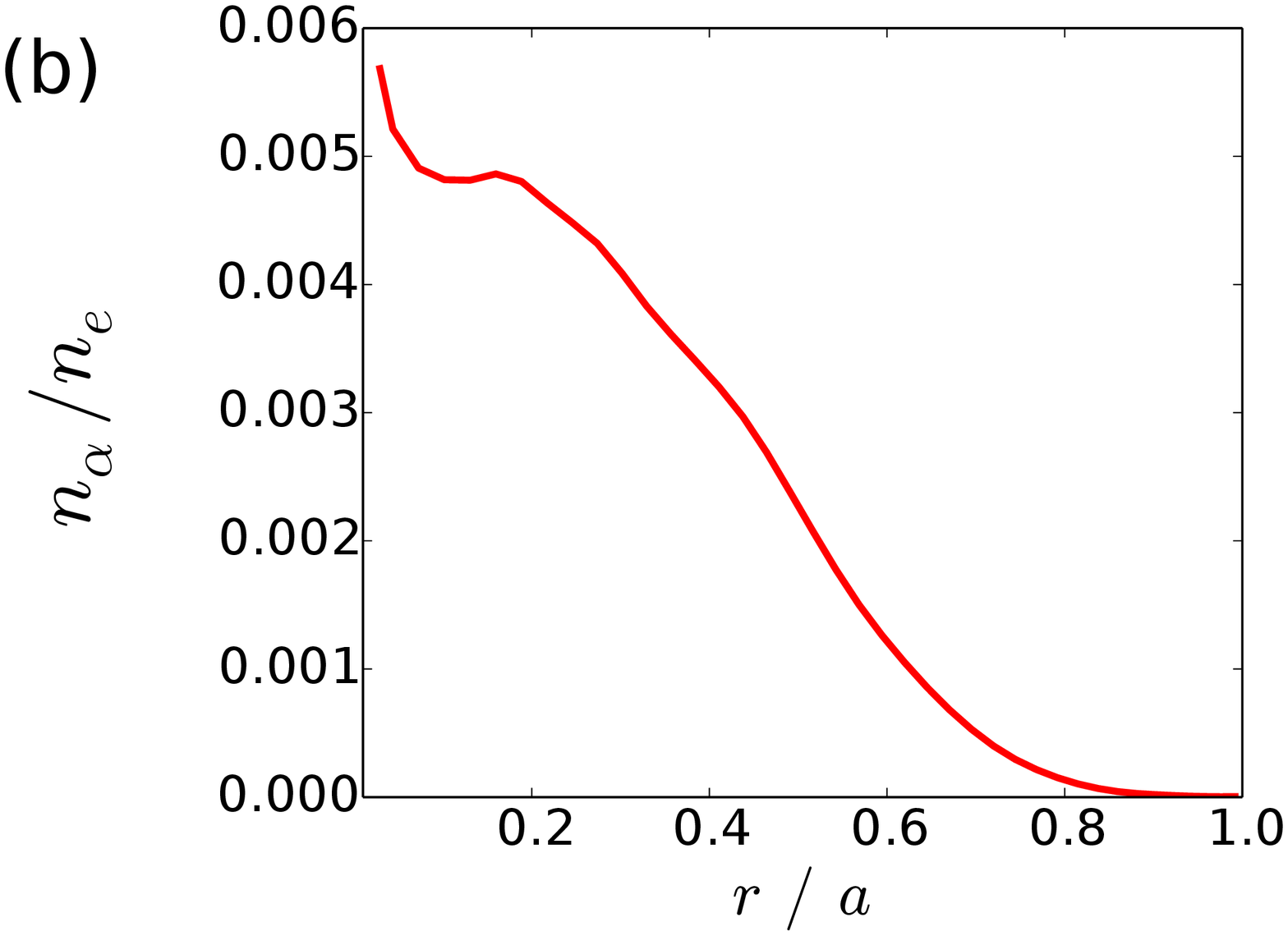}
\includegraphics[width=0.4\textwidth]{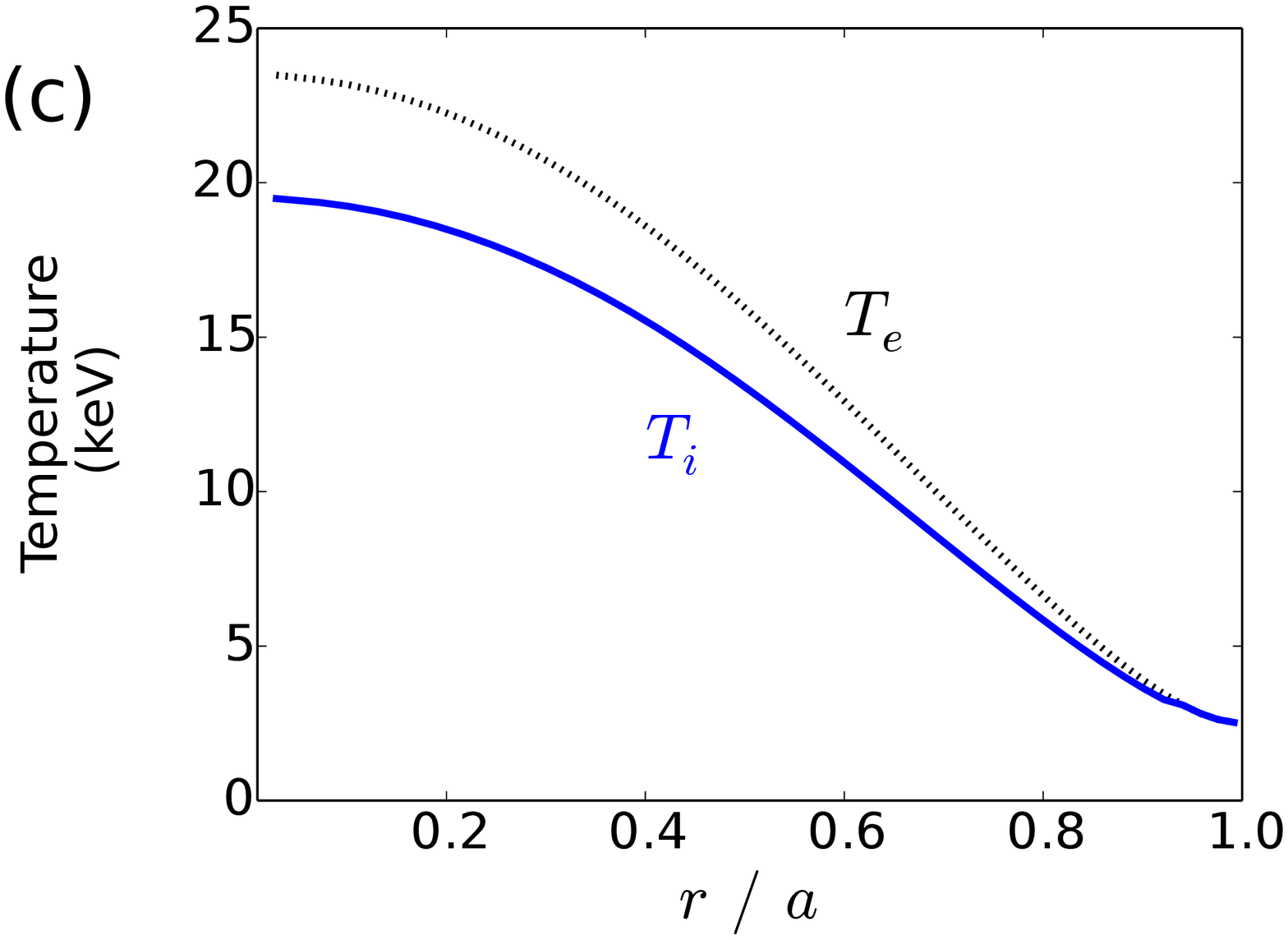}
\includegraphics[width=0.4\textwidth]{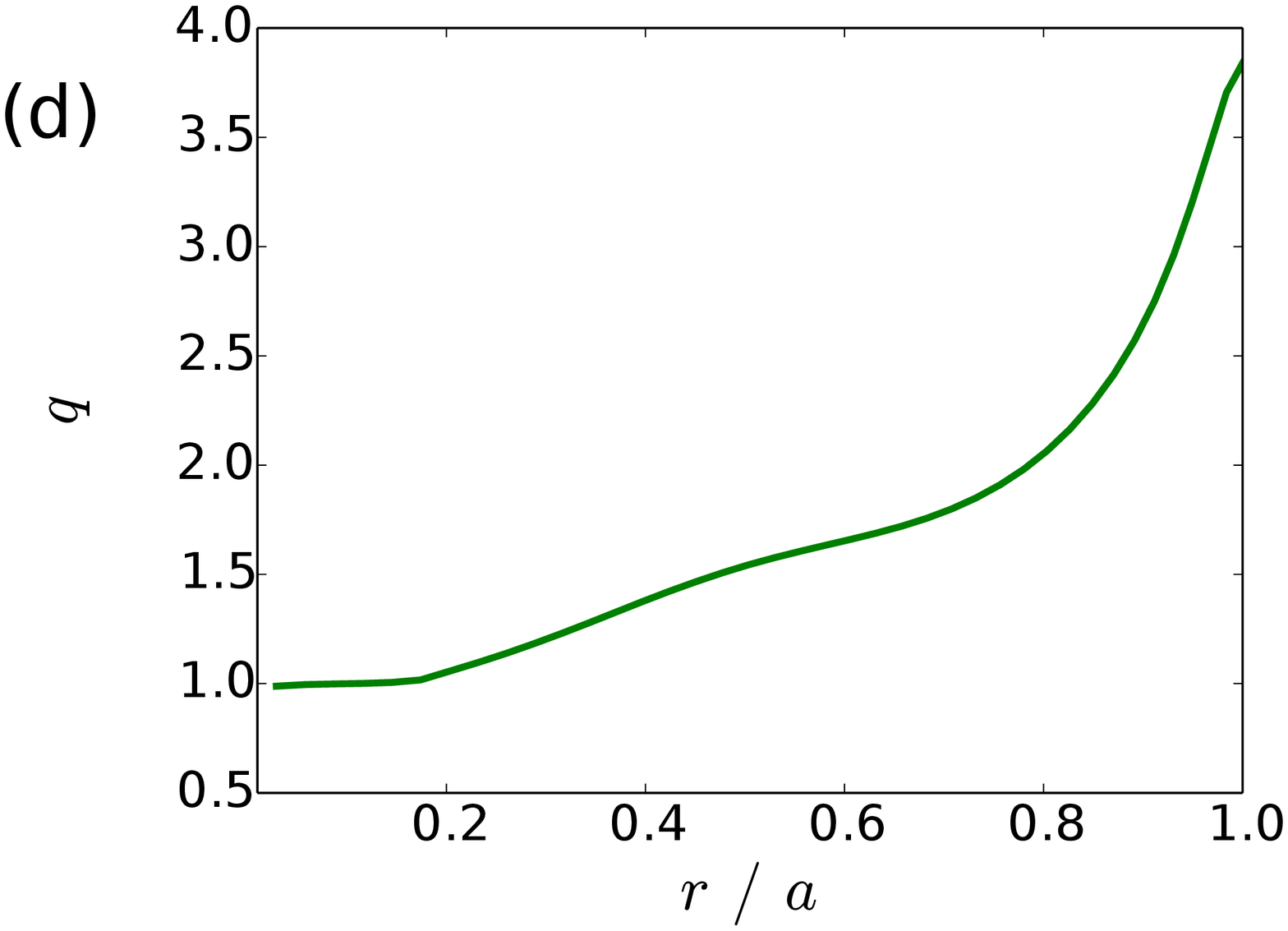}
\caption{\label{iterprofile} Global steady-state profiles for ITER shot 10010100. For these plots, $r$ is the half-diameter on the plane of the magnetic axis, and serves to label the flux surface. Plots show, as a function of radius, (\textit{a}) number density of bulk ions, electrons and cold He ash, (\textit{b}) relative concentration of alpha particles, (\textit{c}) temperature, and (\textit{d}) safety factor $q$ }
\end{center}
\end{figure}

%\begin{table}
%\caption{\label{iterlocal} Local properties used for flux-tube simulations.}
%\begin{center}
%\begin{tabular}{@{}lcccccccc}
%\hline
%Case & & $q$ & $\hat{s}$ & $T_e / T_i$ & $R/L_{n_i}$ & $R/L_{T_i}$ & $R/L_{n_\alpha}$ & $n_\alpha / n_e$  \\
%\hline
%Cyclone base case & & 1.39 & 0.8 & 1.0 & 2.2 & 6.9 & (varies) & (varies) \\
%ITER shot 10010100 & & 1.66 & 0.392 & 1.18 & -1.17 & 7.13 & 24.8 & 0.122\%  \\
%\end{tabular}
%\end{center}
%\end{table}

\begin{figure}
\begin{center}
\includegraphics[width=0.5\textwidth]{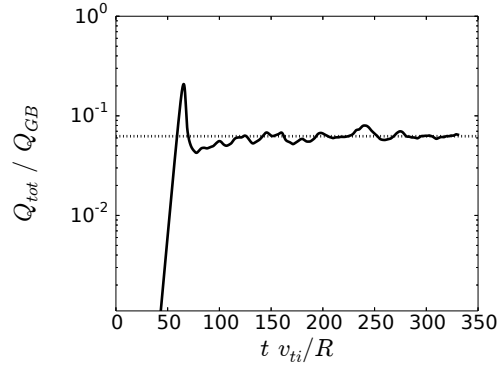}
\caption{\label{iterhflux} The time-evolution of the total heat flux for the ELMy H-mode ITER shot 10010100. }
\end{center}
\end{figure}

We choose to depart from the cyclone base case since it is based on the geometry of DIII-D, a tokamak for which the flux tube approximation is questionable for alpha particles. The test case here is an projected ITER ELMy H-mode scenario (case \#10010100) from the CCFE 2008 public release database \citep{Roach2008,Campbell2001,Budny2002}, and the radial profiles were simulated with the \texttt{PTRANSP} code. Figure \ref{iterprofile} shows radial profiles of some of the equilibrium properties. We will use a Miller expansion of the geometry about a flux tube on the surface defined by $r / a = 0.6$, which gives the following geometrical properties: safety factor $q = 1.66$, magnetic shear $\hat{s}=0.39$, ellipticity $\kappa = 1.53$ (with $a \kappa'(r) = 0.35$), triangularity $\delta = 0.22$ ($a \delta'(r) = 0.41$), and a Shafronov shift derivative of $\Delta'(r) = -0.097$. Electrons were assumed to be adiabatic again, with an alpha particle concentration of $n_\alpha / n_e = 0.12\%$, and an ash concentration of $n_\mathrm{ash} / n_e = 7.9\%$. The gradient length scales were $a/L_{n_e} = 0.0$, $a/L_{n_i} = -0.37$, $a/L_{n_\mathrm{ash}} = 0.95$, and $a/L_{n_\alpha} = 6.9$. The main ions were taken to be a species with an averaged mass weighted by the density of deuterium, tritium, and a small amount of heavy impurities, resulting in $m_i / m_D = 1.484$, and $Z_i = 1$. The ash is assumed to be at the same temperature as the ions: $T_i = 0.847 T_e = 10.9 \mathrm{keV}$. The box size is $318.9 \rho_i \times 157 \rho_i$ in $x$ and $y$ respectively, with $N_x = 96$ and $N_y = 128$. This large box size is not strictly necessary, but is to ensure that many alpha-particle gyroradii fit inside the simulation domain. The parallel and velocity-space resolutions was the same as for cyclone above: $N_\theta = 32$, $N_v = 16$, and $N_\lambda = 33$. The total heat flux resulting from this simulation is show in figure \ref{iterhflux} for reference.

\subsection{Characteristic time scales}

Define the alpha particle transport time as a characteristic timescale on which the turbulent particle flux acts. It is found by balancing the appropriate terms in the transport equation:
\begin{equation}\label{transporttimedef}
\frac{\partial F_{0\alpha}}{\partial t} + \nabla \cdot \boldsymbol{\Gamma}(E) \sim \frac{F_{0\alpha}}{\tau_\Gamma} - \frac{\Gamma(E)}{L_{n_\alpha}},
\end{equation}
where $L_{n_\alpha}$ is chosen as the characteristic length scale on which the alpha particle flux varies. This serves to define:
\begin{equation}\label{taugamdef}
\tau_\Gamma(E) \equiv \frac{L_{n_\alpha} F_{0\alpha}}{ \Gamma(E)}.
\end{equation}

We wish to compare this transport time to a timescale representative of the effects of collisions. The energy-diffusion term is given by:
\begin{equation}\label{collop}
C_E\left[F_{0\alpha}\right] = \sum\limits_{s=i,e} \frac{1}{v^2} \frac{\partial}{\partial v} \left( \nu_{s}^{\alpha s} v^3 F_{0\alpha} + \frac{1}{2} \nu_{\|}^{\alpha s} v^4 \frac{\partial F_{0\alpha}}{\partial v} \right),
\end{equation}
where:
\begin{equation}\label{nusdef}
\nu_s^{\alpha s} = \frac{16 \pi n_s Z_s^2 e^4 \ln \Lambda_{\alpha s}}{m_\alpha} \frac{1}{T_s} \frac{v_\alpha}{v} G \left( \frac{v}{v_{ts}} \right),
\end{equation}
\begin{equation}\label{nupardef}
\nu_\|^{\alpha s} = \frac{16 \pi n_s Z_s^2 e^4 \ln \Lambda_{\alpha s}}{m_\alpha^2 } \frac{2}{v^3} G \left( \frac{v}{v_{ts}} \right),
\end{equation}
and $G$ is the Chandrasekhar function:
\begin{equation}\label{chandraGdef}
G\left(x\right) \equiv \frac{ \text{Erf}(x) - \frac{2 x}{\sqrt{\pi}} e^{-x^2}}{2 x^2}.
\end{equation}
In the limit $v_{ti} \ll v \ll v_{te}$, the collision operator is dominated by the $\nu_{s}^{\alpha e}$ term, and at lower energies, $\nu_s^{\alpha i}$ and $\nu_\|^{\alpha i}$ become more important. For the slowing-down distribution $\partial F_{S\alpha} / \partial v = - 3 v^2 F_{S\alpha} / \left( v_c^3 + v^3 \right)$, so we will use this to estimate the derivative in the $\nu_\|$ terms in equation (\ref{collop}) for a slightly more general $F_{0\alpha}$.

Adding all these terms, we can define a total collision time by:
\begin{equation}\label{taucdef}
\frac{1}{\tau_c} \sim \frac{1}{F_{0\alpha}} C_E\left[F_{0\alpha}\right] \approx \sum\limits_{s=i,e} \frac{16 \pi n_s Z_s^2 e^4 \ln\Lambda_{\alpha s}}{m_\alpha^2} \left[ \frac{m_\alpha  }{T_s v} - \frac{3}{v_c^3 + v^3}  \right]G\left( \frac{v}{v_{ts}} \right) .
\end{equation}

\begin{figure}
\begin{center}
\includegraphics[width=0.7\textwidth]{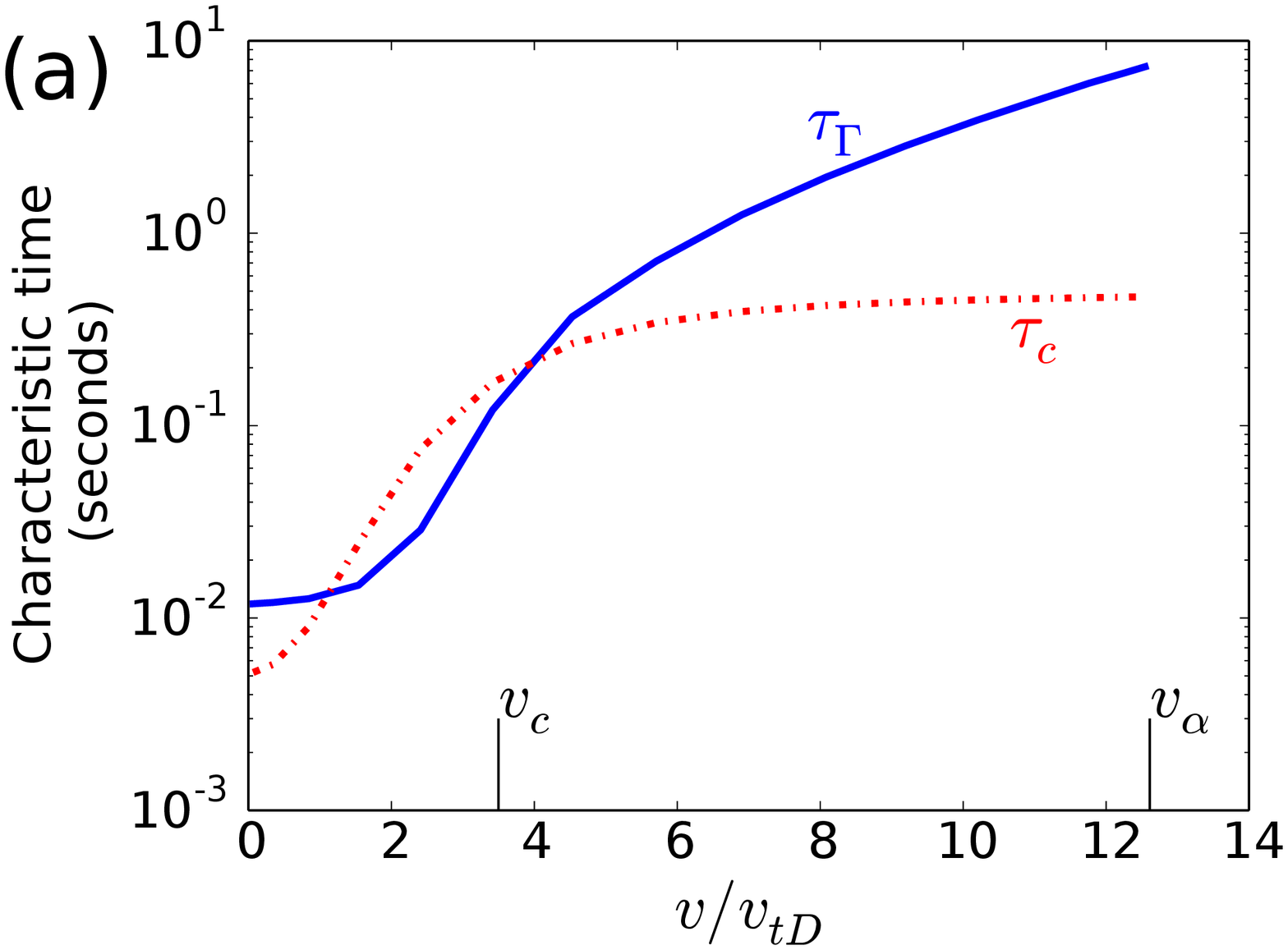}
\includegraphics[width=0.7\textwidth]{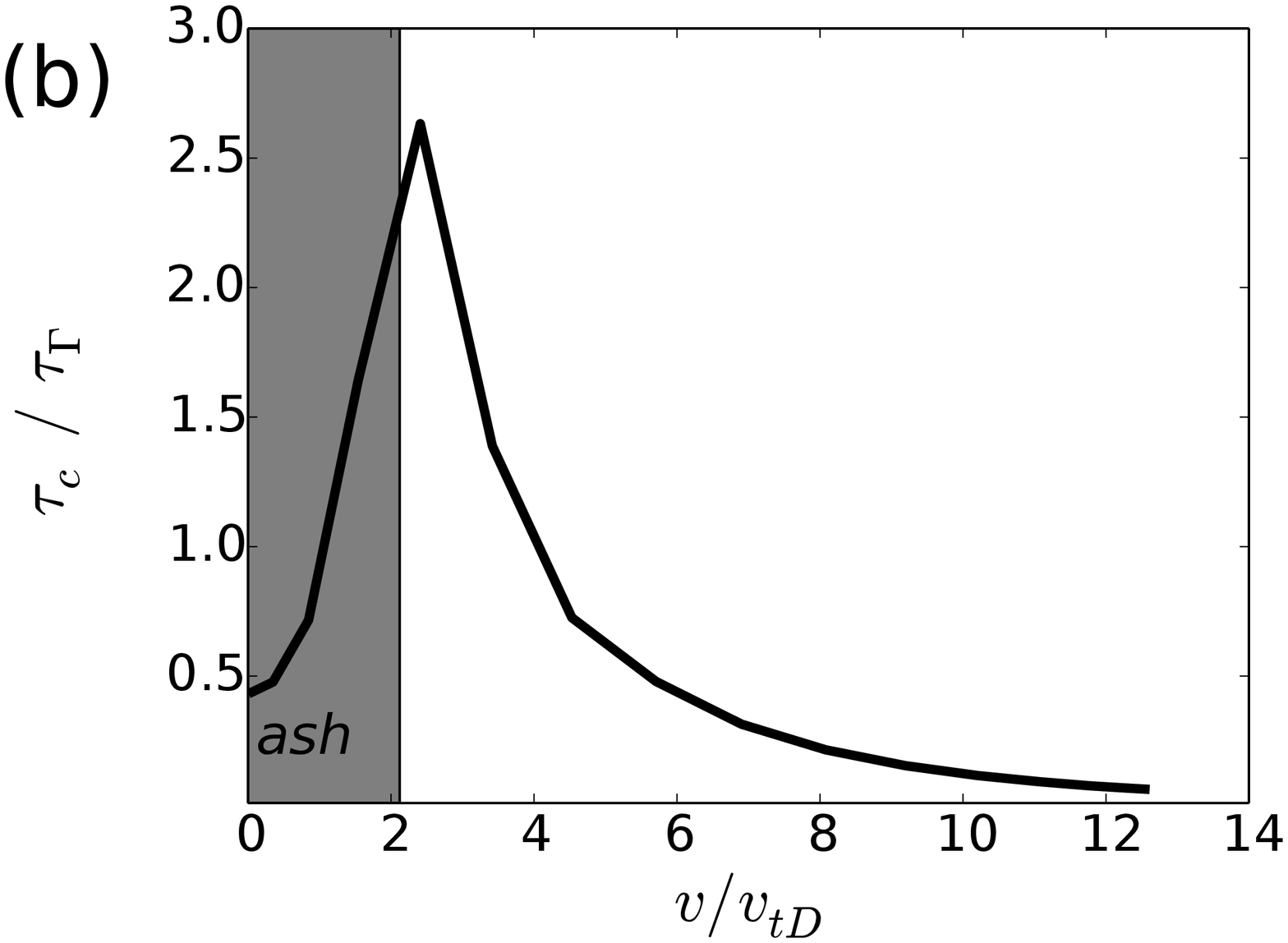}
\caption{\label{conftime} Comparing characteristic transport time ($\tau_\Gamma$) and collision time ($\tau_c$) as a function of velocity. (\textit{a}) shows the characteristic times directly, whereas (\textit{b}) shows the relative importance of radial transport compared to collisions: $\tau_c / \tau_\Gamma$. The shaded region represents the area up to $v = 3 \times v_{t,\text{ash}}$ (assuming $T_\text{ash} = T_i$), where the slowing-down distribution is not valid and is dominated by Maxwellianised Helium ash.}
\end{center}
\end{figure}

Now, by comparing equations (\ref{taugamdef}) and (\ref{taucdef}), we can make a reasonable estimate of how relevant a transport term would be in an equation like (\ref{transporteqn}). This is shown in figure \ref{conftime}. Around $E_\alpha = 3.5$ MeV, we see that collisions are dominant over transport, and $\tau_c$ flattens out to the slowing-down time $\tau_s$ as expected. Also, if $\tau_\Gamma$ is interpreted as a particle confinement time, we expect hot alphas to be well-confined on the order of several seconds. This is roughly consistent with previous work \citep[see][]{Angioni2009}. The actual alpha particle confinement time (defined as the average number of alpha particles leaving a flux surface divided by the total number of particles contained within the flux surface) is about $2.8$s in our simulation, consistent with the energy confinement time estimated in table 1.

However, at lower energies, but still well above the ion or ash temperatures, the radial transport of alpha particles becomes important compared to collisions. For this case, it can be seen in figure \ref{conftime}(\textit{b}) that the relative importance peaks near the critical speed, but there is no reason to believe this is more than coincidence: there are a number of parameters that could, in principle, be independently tuned. For example, the transport time scale scales quadratically with both $\rho^*$ and $L_{n\alpha}$, neither of which would have a direct effect on the characteristic collision time.

\section{Summary and discussion}

In this work, we have laid out our framework for employing gyrokinetics to study the behavior of isotropic non-Maxwellian fast ions. From first principles, we critically analyze several common assumptions made in the analysis of hot alpha particles. The basic conclusion is that, electrostatically, alpha particles react only passively to ion-scale turbulence. Even at high charge densities of around $20\%$ of the total charge, the effect is simply that of diluting the drive from the main ions. 

On the other hand, if one wishes to analyze in what manner are the alpha particles are advected by the background turbulence, it is clear even from linear theory that the correct distribution function must be used. The reason for this is because the perturbed distribution function depends linearly on the radial gradient of the equilibrium distribution. If an incorrect model distribution is used (e.g. the ``equivalent Maxwellian''), applying the correct velocity-space dependence of $\nabla F_{S\alpha}$ either presupposes the relevant region of velocity space (e.g. by employing a linear fit to figure \ref{gradF0vsE} near $E=0$), or is not based on physical principles. At least three major gyrokinetic codes (\texttt{GS2}, \texttt{GENE}, and \texttt{GYRO}) have the capability to model non-Maxwellian species, so there is little reason to continue using an inadequate model. Previous effort spent by other groups in the analysis of alpha particles using the equivalent Maxwellian is certainly not wasted, however. As long as alpha particles are trace, a direct conversion between the two is possible and was presented.

Armed with a good analytical estimate of the equilibrium distribution at high energy: the classical slowing-down distribution, we proceed to analyze the strength of advection relative to collisionality as a function of energy. Where the ratio of these characteristic scales approaches or exceeds unity, the slowing-down distribution itself is expected to be wrong. Our analysis of a projected ITER shot shows that this is indeed the case. This means that, although alpha particles are predicted to be well-confined overall (consistent with experiment, see \citet{Pace2013}), energy-dependent transport is strong enough to affect the slow time evolution of the equilibrium distribution function, at least at some energies for some combination of reasonable parameters. The very high-energy part of the slowing-down tail near $3.5 \mathrm{MeV}$ remains relatively unaffected by the turbulence, as expected. 

It should be kept in mind that ours is a local flux-tube analysis. Depending on the tokamak, shot, and radius, the gradient length scales can change significantly over the orbit size of a high-energy alpha particle. By restricting ourselves to ITER shots or significantly uniform profiles, we strive to alleviate this concern. It is believed that the primary results presented herein are robust to these caveats, but additional study of these effects is welcome.

Our results clearly suggest that a transport study capturing the coupled radial and energy dependence of the alpha particle distribution is warranted. By using a passive tracer model with transport coefficients found from a series of fully nonlinear simulations, one can evolve $F_{0\alpha} \left( v, r \right)$. It is this distribution which ought to be used locally to obtain more confident results for the behavior of alpha particles, especially for finite-$\beta$ simulations in which alpha particles are expected to play a far more active role.

\section*{Acknowledgements}

The authors would like to thank M. Barnes, F. Parra, G. Hammett, and A. Schekochihin for their ideas and insightful discussions, G. Colyer for the use of computer time on Helios, and T. Fredian for his help in accessing the CCFE public tokamak profile database. The Wolfgang Pauli Institute in Vienna and CIEMAT in Madrid have graciously hosted workshops and meetings at which much of this work was inspired and performed. The use of the Helios (IFERC) and Edison (NERSC) supercomputers have been critical to the results presented here. This material is based upon work supported by the U.S. Department of Energy, Office of Science, Office of Fusion Energy Science, under award numbers DEFG0293ER54197 and DEFC0208ER54964. 

%\appendix

\bibliographystyle{jpp}
\bibliography{mylib}

\end{document}